\documentclass[<1p>]{elsarticle} %authoryear

\usepackage{amsmath, amsthm, amssymb, amsthm}
\usepackage{subcaption}
\usepackage[flushleft]{threeparttable}
\usepackage{tabu,multirow}
\usepackage{color}
\usepackage{etoolbox}
\usepackage{float}
\usepackage{booktabs}

\graphicspath{ {Pictures/} }

\begin{document}
\begin{frontmatter}
\begin{abstract}
The properties of Maximum Likelihood estimator in mixed causal and noncausal models with a generalized Student’s $t$ error process are reviewed. 
Several known existing methods are typically not applicable in the heavy-tailed framework. To this end, a new approach to make inference on causal 
and noncausal parameters in finite sample sizes is proposed. It exploits the empirical variance of the generalized Student's $t$, without the existence of 
population variance. Monte Carlo simulations show a good performance of the new variance construction for fat tail series. Finally, different existing 
approaches are compared using three empirical applications: the variation of daily COVID-19 deaths in Belgium, the monthly wheat prices, and the monthly 
inflation rate in Brazil.
\end{abstract}
\begin{keyword}
MLE \sep noncausal models \sep generalized Student's t-distribution \sep robust inference.\\
\emph{JEL:} C22
\end{keyword}
\title{Inference in mixed causal and noncausal models with generalized Student's t-distributions}
\author[add1]{Francesco Giancaterini\corref{cor1}}
\ead{f.giancaterini@maastrichtuniversity.nl}
\cortext[cor1]{Corresponding author}
\author[add1]{Alain Hecq}
\address[add1]{Maastricht University, Department of Quantitative Economics, School of Business and Economics, P.O.box 616, 6200 MD, Maastricht, The Netherlands.}
\end{frontmatter}

\section{ Introduction}
Mixed causal and noncausal models (MARs) are time series processes that consist of lead and lag components. Unlike causal 
models (e.g. ARMA models), these specifications can capture nonlinear features, such as bubbles, which can be defined as economic or financial
processes that undergo a rapid increase followed by a sudden crash. MAR has manifested 
its values in various applications, for instance, commodity prices, inflation rates, bitcoin, and other forms of equity (see inter alia \cite{hencic2015noncausal},
\cite{gourieroux2016filtering}, \cite{bec2020mixed}, \cite{karapanagiotidis2014dynamic}, \cite{lof2017noncausality}, 
\cite{hecq2020selecting}, \cite{gourieroux2020convolution}, and \cite{gourieroux2021forecast}). Furthermore, the interpretation of MARs
is rooted in economic theory. More specifically, these models can be interpreted as conditions in which economic agents have more 
information than econometricians. For this reason, mixed causal and noncausal models are linked to the existence of non-fundamentalness 
in structural econometric models (see \cite{alessi2011non}, and \cite{lanne2013noncausal}). Finally, as MARs can be utilized to test for the time 
reversibility of a stochastic process, they can also be employed to detect business cycle asymmetry. In recent decades seminal work has sparked a 
renewed interest in the topic (see \cite{breidt1992time}, \cite{ramsey1996time}, and \cite{proietti2020peaks}).
Despite the wide applicability of MARs, their estimation and inference are far from trivial.\\
\indent A common occurrence in the MAR literature is the assumption of an error term that is distributed according to a generalized Student's $t$
in order to capture the heavy tails in relevant data. Typically, the degrees of freedom of the Student's $t$ are initially unknown and must
be estimated. Many studies have revealed that the estimated degrees of freedom of the generalized Student’s $t$
lie between 1.5 and 2 (see \cite{fries2019mixed}, \cite{hecq2019predicting}, \cite{hecq2021forecasting}, and \cite{hecq2020mixed}).
This empirical finding imposes a difficulty for inference in MARs, namely the variance does not exist. In order to bypass this difficulty,
certain methods have been proposed: (1) using a different asymptotic framework (\cite{davis1985limit}); (2) adopting other distributions
(\cite{fries2019mixed}, and \cite{fries2021conditional}); (3) employing bootstrap estimators (\cite{cavaliere2020bootstrapping}, in the case of purely 
noncausal models). Our novel strategy involves the use of simulations to identify a robust estimator of the variance which performs optimally when 
dealing with small sample sizes. An estimator of this kind facilitates the inference of MARs by using a standard $t$-test. Extensive Monte Carlo simulations %controlla trattino del t test
show that the $t$-tests have empirical rejection frequencies (E.R.F.) close to the nominal significance level.\\
\indent The rest of the paper is organized as follows; Section 2 introduces mixed causal and noncausal models. It presents the 
different ways of obtaining the expected Fisher Information matrix for MARs and the existing strategies are briefly reviewed.
Section 3 proposes a new approach to compute the standard errors of causal and noncausal parameters based on a robust 
estimator of the residuals. In Section 4, Monte Carlo simulations are used to compare the performance of current methodologies with 
the novel approach proposed here. Section 5 is dedicated to demonstrating the empirical applications on three different time series. Section 6 concludes.
 
\section{Mixed causal and noncausal models}
Consider the univariate MAR($r$,$s$) defined as:
\begin{equation}
\phi (L)\varphi (L^{-1})y_{t}=\epsilon _{t},
\end{equation}%
where $L$ is the backshift operator and $L^{-1}$ produces leads such that $L^{-1}y_{t}=y_{t+1}$. In order to have an easier notation, 
we assume that $y$ has been demeaned. We denote $\phi (L)$ as causal/autoregressive polynomial of order $r$, $\varphi (L^{-1})$ as 
noncausal/lead polynomial of order $s$, and $p=r+s$. It is assumed that both the polynomials $\phi(z)$ and $\varphi(z)$ have their roots
outside the unit circle:
\begin{equation}
\phi(z) \neq 0 \  \ and \  \ \varphi(z) \neq 0 \ \ for \ \ |z| \leq 1.
\end{equation}
Note that purely causal or noncausal models are obtained by respectively setting $\varphi (L^{-1})=1$ and $\phi (L)=1$.
Furthermore, $\epsilon _{t}$ is an independent and identically ($i.i.d.$) generalized Student's $t$ sequence of random variables (the non-Gaussianity
assumption of the error term is required to identify noncausal  from causal models) with a mean of zero and finite variance ($\sigma^{2}$), such that its density function is:
\begin{equation}
f(\epsilon_{t},\nu, \eta)= \frac{\Gamma(\frac{\nu+1}{2})}{\Gamma(\frac{\nu}{2}) %
\sqrt{\pi \nu} \eta} \Biggl[ 1 + \frac{1}{\nu} \biggl(\frac{\epsilon_{t}}{ \eta} \biggr)^{2}  \Biggr]^{\frac{-(\nu + 1)}{2}},
\end{equation}
with corresponding (approximate) log-likelihood function:
\begin{multline}
l(\boldsymbol{\phi}, \boldsymbol{\varphi}, \nu , \eta)= (T-p) \Biggl[ \ln \biggl( \Gamma \biggl( \frac{\nu+1}{2} \biggr) \biggr) - %
\ln \biggl( \sqrt{\nu \pi} \eta \biggr) - \ln \biggl( \Gamma \biggl( \frac{\nu}{2}  \biggr) \biggr) \Biggr] + \\ 
- \frac{\nu+1}{2} \sum \limits_{t=r+1}^{T-s} \ln \biggl( 1 + \frac{1}{\nu} \biggl( \frac{ \phi(L) \varphi(L^{-1})y_{t} }{ \eta}  \biggr) ^{2} \biggr),
\end{multline}
where the vectors $\boldsymbol{\phi}=(\phi _{1},...,\phi _{r})$ and $\boldsymbol{\varphi }=(\varphi_{1},...,\varphi_{s})$ respectively collect the causal and 
the noncausal coefficients. In addition, $\nu$ denotes the degrees of freedom, $\eta$ is the scale parameter and:
\begin{equation*}
\sigma^{2}=\eta^{2} \frac{\nu}{\nu - 2},
\end{equation*}
is the variance of the error term. Since the first $r$ and the last $s$ observations are lost in the MAR($r$,$s$) estimation, in (1), the parameter vectors $\boldsymbol{\phi}$ 
and $\boldsymbol{\varphi }$ can be estimated by an AMLE approach. AMLE refers to the approximate Maximum Likelihood estimators.\\
\indent \cite{lanne2011noncausal} employ a more general assumption than we do. More specifically, they use mixed causal and noncausal 
models which are characterized by an innovation term whose density function satisfies conditions (A1)-(A7) of \cite{andrews2006maximum}. The
generalized Student's $t$ is one of them. In order to present their theorem, some provisional notation is necessary.\\
\indent Let $\zeta_{t} \sim i.i.d \ (0,1)$ and define the AR($r$) stationary process $u_{t}^{*}$ by $\phi_{0}(L)u_{t}^{*}=\zeta_{t}$ and the AR($s$) stationary process 
$v_{t}^{*}$ by $\varphi_{0}(L)v_{t}^{*}=\zeta_{t}$. Let also define $U_{t-1}^{*}=(u_{t-1}^{*},\dots ,u_{t-r}^{*})$, $V_{t-1}^{*}=(v_{t-1}^{*},\dots ,v_{t-s}^{*})$ 
and the associated covariance matrices $\Gamma_{U^{*}}=Cov(U_{t-1}^{*})$, $\Gamma_{V^{*}}%
=Cov(V_{t-1}^{*})$ and $\Gamma_{U^{*} V^{*} }=Cov(U_{t-1}^{*},V_{t-1}^{*})=\Gamma_{V^{*} U^{*}}^{\prime}$.\\
\newline
\textbf{Theorem 1} (by \cite{lanne2011noncausal}) Given conditions (A1)-(A7) of Andrews et al. (2006), 
there exists a sequence of local maximizers $\widehat{\boldsymbol{\theta }}=(\widehat{\boldsymbol{\phi }},\widehat{\boldsymbol{\varphi }},\widehat{%
\eta },\widehat{v}$) of $l_{t}(\boldsymbol{\theta})$ in (4) such that%
\begin{equation*}
(T-p)^{1/2}(\widehat{\boldsymbol{\theta }}-\boldsymbol{\theta }_{0})%
\xrightarrow{d}N(0, \mathcal{I(\boldsymbol{\theta})}^{-1}),
\end{equation*}
where:
\begin{equation}
\mathcal{I(\boldsymbol{\theta})}=-E\bigl[\frac{\delta ^{2}l(\boldsymbol{%
\theta })}{\delta \boldsymbol{\theta}\delta \boldsymbol{\theta}^{\prime }}%
\bigr]=diag(\Sigma,\Omega),
\end{equation}
is the expected Fisher information matrix (it yields the Cramér–Rao bound, which gives the lower bound of the 
asymptotic covariance matrix of the ML estimators). The $diag$ notation in (5) implies that the expected Fisher Information matrix is asymptotically a block diagonal matrix.
In other words, the two blocks $\Sigma$ and $\Omega$ are asymptotically independent such that they can be treated separately.
$\Sigma$ is the expected Fisher Information matrix of the AML estimators $(\boldsymbol{\phi} ,\boldsymbol{\varphi})$, defined 
by \cite{lanne2011noncausal} as:
\begin{equation}
\Sigma =  \begin{bmatrix} \mathcal{J} \Gamma_{U^{*}} & \Gamma_{U^{*} V^{*} } & \\%
\Gamma_{V^{*} U^{*}} & \mathcal{J} \Gamma_{V^{*}}\end{bmatrix}
=
 \begin{bmatrix}\sigma^{2}\tilde{ \mathcal{J}} \Gamma_{U^{*}} & \Gamma_{U^{*} V^{*} } & \\%
\Gamma_{V^{*} U^{*}} &\sigma^{2}\tilde{ \mathcal{J}} \Gamma_{V^{*}}\end{bmatrix},
\end{equation}
where:
\begin{equation}
\mathcal{J}=\sigma^{2}\tilde{ \mathcal{J}}= \int \frac{f'(\epsilon_{t}, \nu, \eta)^{2}}{ f(\epsilon_{t},\nu,\eta)} d\epsilon_{t}.
\end{equation}
On the other hand, $\Omega $ is the expected Fisher Information matrix of the distributional parameters ($\nu$ and $\eta$).
This paper only focuses on matrix $\Sigma$ since we are exclusively interested in the standard errors of causal and noncausal parameters.
It is positive definite if $\mathcal{J}>1$ (see condition (A5) of \cite{andrews2006maximum}), and when the innovation term is distributed 
according to a generalized Student's $t$-distribution, \cite{andrews2006maximum} show that:
\begin{equation*}
\mathcal{J}=\frac{\nu (\nu +1)}{(\nu -2)(\nu +3)}.
\end{equation*}
Hence, $\Sigma$ is positive definite for $\nu > 2$. The shortcoming of the approach proposed by \cite{lanne2011noncausal} is that we cannot consider processes 
characterized by an error term as distributed as a generalized Student's $t$ in cases where the degrees of freedom are less than 2. It is restrictive for heavy-tailed 
time series such as stock and commodity prices, bitcoin, and other form of equity, in cases where the degrees of freedom range between 1.3 and 1.9 (without reaching the 
Cauchy for $\nu =1$ though).\\
\indent \cite{hecq2016identification} introduce an approximative and more straightforward way to compute the matrix defined in (6). This methodology is implemented 
in the R package MARX, and has been applied in several studies. It can be used whenever the error term is $i.i.d.$ and has a density function as expressed in (3). Adapting 
the results obtained by \cite{fonseca2008objective} in the context of MARs, \cite{hecq2016identification} derive the following expected Fisher Information matrix
of causal and noncausal parameters:
\begin{equation}
\Sigma_{D}=-E%
\begin{bmatrix}
\frac{\delta ^{2}l(\boldsymbol{\phi},\eta,\nu | \boldsymbol{\varphi})}{\delta \boldsymbol{\phi}  \boldsymbol{\phi}\prime} & 0 \\ 
0 & \frac{\delta ^{2}l(\boldsymbol{\varphi},\eta ,\nu | \boldsymbol{\phi})}{\delta \boldsymbol{\varphi} \boldsymbol{\varphi}\prime}
\end{bmatrix}.
\end{equation}
Whenever the error term has a finite variance, matrix (8) leads to the same diagonal blocks as in matrix (6).\\
For $\nu <2$, the approximate expected Fisher Information matrix of the causal and noncausal parameters ($\Sigma_{D}$) is still positive definite and provides 
standard errors, unlike $\Sigma$.
However, a closed-form solution for the limiting distribution of the MAR parameters 
does not exist in this context (see \cite{davis1992m}, and \cite{andrews2013model}). This problem could be overcome by employing bootstrapping and simulation-based models.

\section{A new robust estimator}
In this section, we propose a new methodology to compute the standard errors of MAR parameters. Its use is valid with finite sample sizes in instances where the error term is assumed 
to be distributed according to a generalized Student’s t-distribution. The next section will use Monte Carlo simulations to empirically show that our new estimator performs optimally for 
$\nu \in (1,D]$, with $D<\infty$. This is true, although it is not possible to derive the theoretical limiting distributions of these parameters in the heavy-tail framework.\\ 
\indent In Section 2, it is stated that the variance of the error term ($\sigma^{2}$) multiplies the block diagonal matrices of $\Sigma$. Since the generalized
Student's $t-$distribution with heavy-tailed innovations is characterized by infinite variance, the expected Fisher Information matrix  cannot be computed in this context.
Our alternative strategy consists of replacing the error term's variance with the variance of residuals ($\sigma_{\hat{\epsilon}}^{2}$) in (6). Furthermore, we expect the 
residuals to have a wide range of values, especially in cases where the population variance is infinite. In order to decrease the effect of huge outliers, we estimate the 
standard deviation of the residuals using the robust estimator introduced by \cite{rousseeuw1993alternatives}:
\begin{equation}
\sigma_{\hat{\epsilon}}=k \times MAD_{\hat{\epsilon}}.
\end{equation}
where $MAD$, also known as median absolute deviation, is a robust estimator of the variability of residuals:
\begin{equation}
MAD_{\hat{\epsilon}}=median(|\hat{\epsilon}_{i} - median(\hat{\epsilon}_{i})| ),
\end{equation}
and $k$ is a scalar value. \cite{rousseeuw1993alternatives} state that the value of $k$ depends on the distribution of residuals, and that in the Gaussianity case, $k = 1.48$ ensures a robust
estimate of the standard deviation. To detect $k = 1.48$ through an empirical approach, we apply a Monte Carlo experiment where a Gaussian error term is simulated (with an expected value equal 
to 0 and standard deviation equal to 5), considering $T = 1000$ observations and $N = 700000$ replications. A large number of replications is required to obtain an empirical density function as 
accurately as possible. In each replication, we compute the value of $k$ using equation (9). In this way, the experiment yields as many estimates of $k$ as the number of replications. To analyze 
the behavior of these estimates, we compute the empirical density function of $k$ using the kernel density estimation. It is well known that extreme values of $k$ can affect the non-parametric estimation 
(see \cite{kim2012robust}). To this end, we only consider the values of $k$ that lie in the interval
\begin{equation}
\bigl[Q1-3 \times IQR, \ Q3+3 \times IQR\bigr],
\end{equation}
in the estimation procedure. Note that $Q1$ and $Q3$ indicate the first and the third quartile of $k$ respectively, and $IQR$ is its interquartile range. Figure 1 shows the result obtained: a distribution 
centered around the mode 1.48, the same value identified by \cite{rousseeuw1993alternatives}.\\
\begin{figure}[H]
\centering
   	 \begin{subfigure}{.4\textwidth}
		\includegraphics[width=\linewidth]{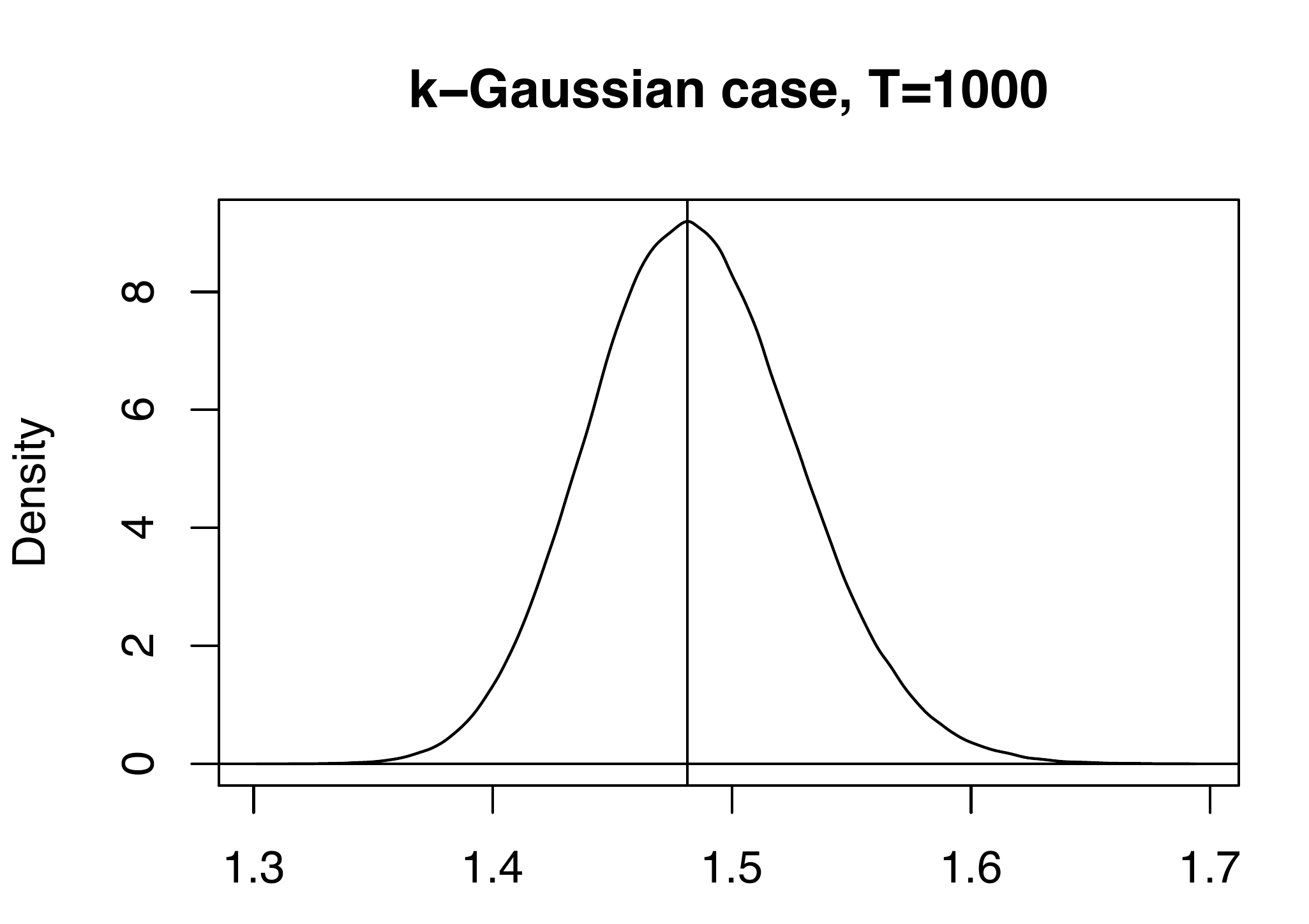}
	\end{subfigure}
	\caption{The empirical density function of $k$ under the assumption of Gaussianity of the residuals is shown. $N=700000$ replications and $T=1000$ observations are considered.}
 \end{figure}
\indent 
Let us now find the value of $k$ that provides a robust estimate of the standard deviation of residuals, under the assumption that the error process follows a generalized Student's $t-$distribution, 
for $\nu \in (1,D]$. In this case, we cannot use $k=1.48$ because, as previously stated, it changes according to the distribution considered. Hence, we apply the same empirical approach
used in the Gaussian case to detect its value. More specifically, after estimating the degrees of freedom $(\hat{\nu})$ of our MAR process, we apply a Monte Carlo experiment where the error term is simulated 
setting $\nu_{0}=\hat{\nu}$ and the sample size is equal to the number of observations detected ($T$). The next step is to take the values of $k$ that lie inside the interval (11), and compute its empirical 
density function. Finally, as in the Gaussian case, we only need to extract a single value of $k$ from its empirical density function. We take its mode and denote it as $k^{*}$. \\
\indent It is important to note that under the assumption of Student's $t$, the standard deviation of the residuals depends on two different parameters: the degrees of freedom ($\nu$) and the sample size ($T$). 
This implies that we can rewrite (9) as:
\begin{equation}
k (\nu, T)=\frac{\sigma_{\hat{\epsilon}} (\nu, T)}{MAD_{\hat{\epsilon}}}.
\end{equation}
In other words, $k$ is a random variable with different density functions depending on different values of $\nu$ and $T$. To investigate how $k$ changes with $\nu$ and $T,$ 
two different Monte Carlo simulations are carried out. In the first one, we analyze how $k$ correspondingly changes for different values of $\nu$, keeping $T$ as fixed. In particular, we consider 
an experiment where the error term is simulated setting $T$=500, $\nu_{0}=(1.2, 1.8, 3, 50, 1000)$ and $N=700000$ replications. The empirical density functions obtained are 
shown in Figure 2. The graphs show how the empirical density functions differ according to the different values of $\nu$.\\ 
\indent The second Monte Carlo experiment analyzes how the empirical distributions of $k$ change with samples of various sizes. The data generating process is now characterized by 
several sample sizes $T=(100, 200, 500, 1000, 3000)$ and degrees of freedom fixed to $\nu_{0}=1.5$. Figure 3 illustrates how the mode of the empirical distributions 
shifts toward the right as $T$ increases. This implies that $k^{*}$ can also be expressed as a function of $\nu$ and $T$. Appendix A provides its value for different $\nu$ and $T.$\\
\indent In conclusion, this approach leads us to the following Fisher Information matrix of the causal and noncausal coefficients:
\newline
\begin{equation}
\Sigma_{R} = \begin{bmatrix} %
\sigma_{\hat{\epsilon}}^{2}\tilde{\mathcal{J}} \Gamma_{U^{*}} & \Gamma_{U^{*} V^{*} }& \\ \Gamma_{%
V^{*} U^{*}} &\sigma_{\hat{\epsilon}}^{2}\tilde{\mathcal{J}} \Gamma_{V^{*}}\end{bmatrix},
\end{equation}
\newline
where:
\begin{equation*}
\sigma_{\hat{\epsilon}}= k^{*}(\nu, T) \times MAD_{\hat{\epsilon}}.
\end{equation*}
\begin{figure}[H]
\centering
    \begin{subfigure}{.26\textwidth}
        \centering
        $k(\nu=1.2, T=500)$
        \includegraphics[width=\linewidth]{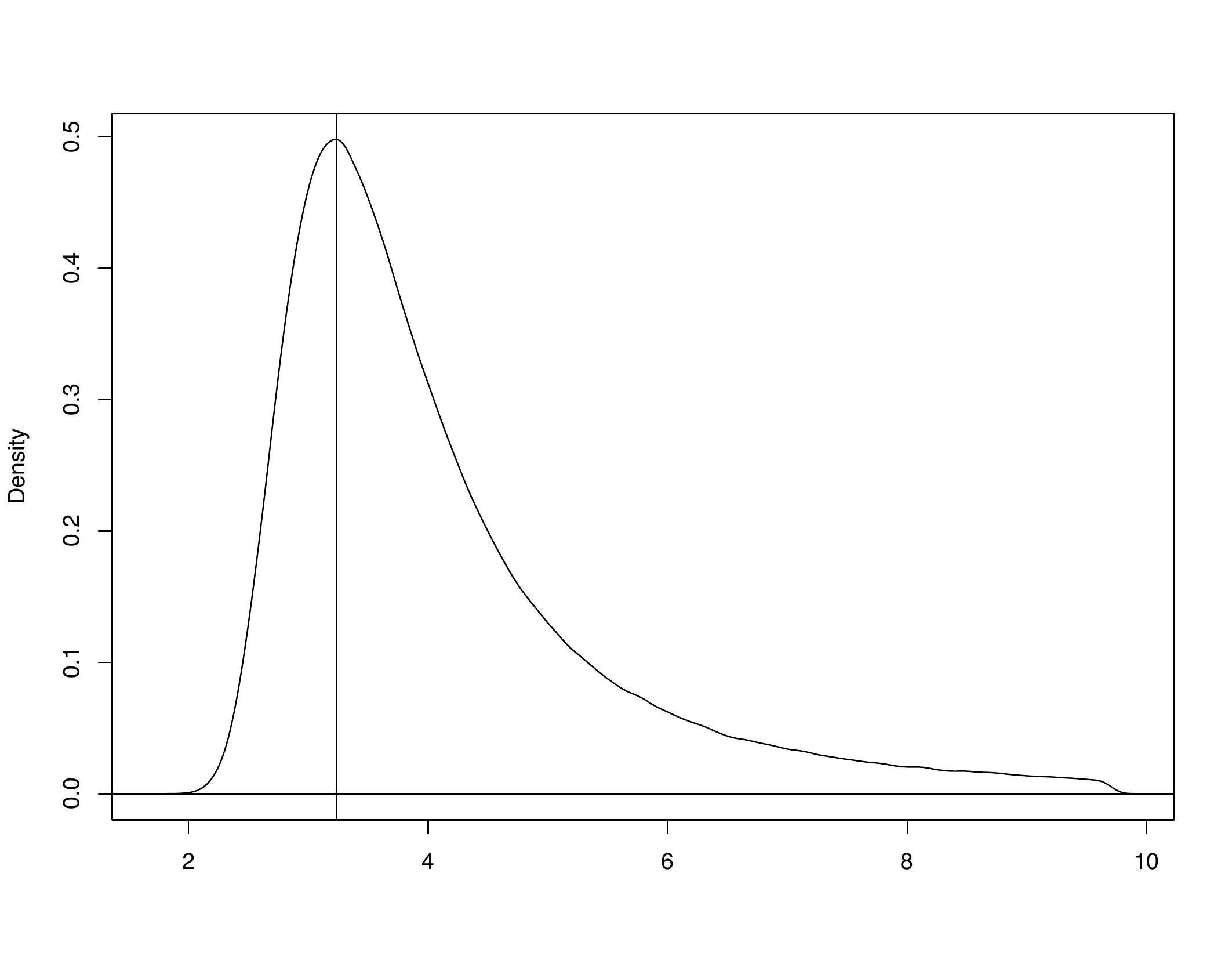}
    \end{subfigure}
    \begin{subfigure}{.26\textwidth}
        \centering
        $k(\nu=1.8, T=500)$
        \includegraphics[width=\linewidth]{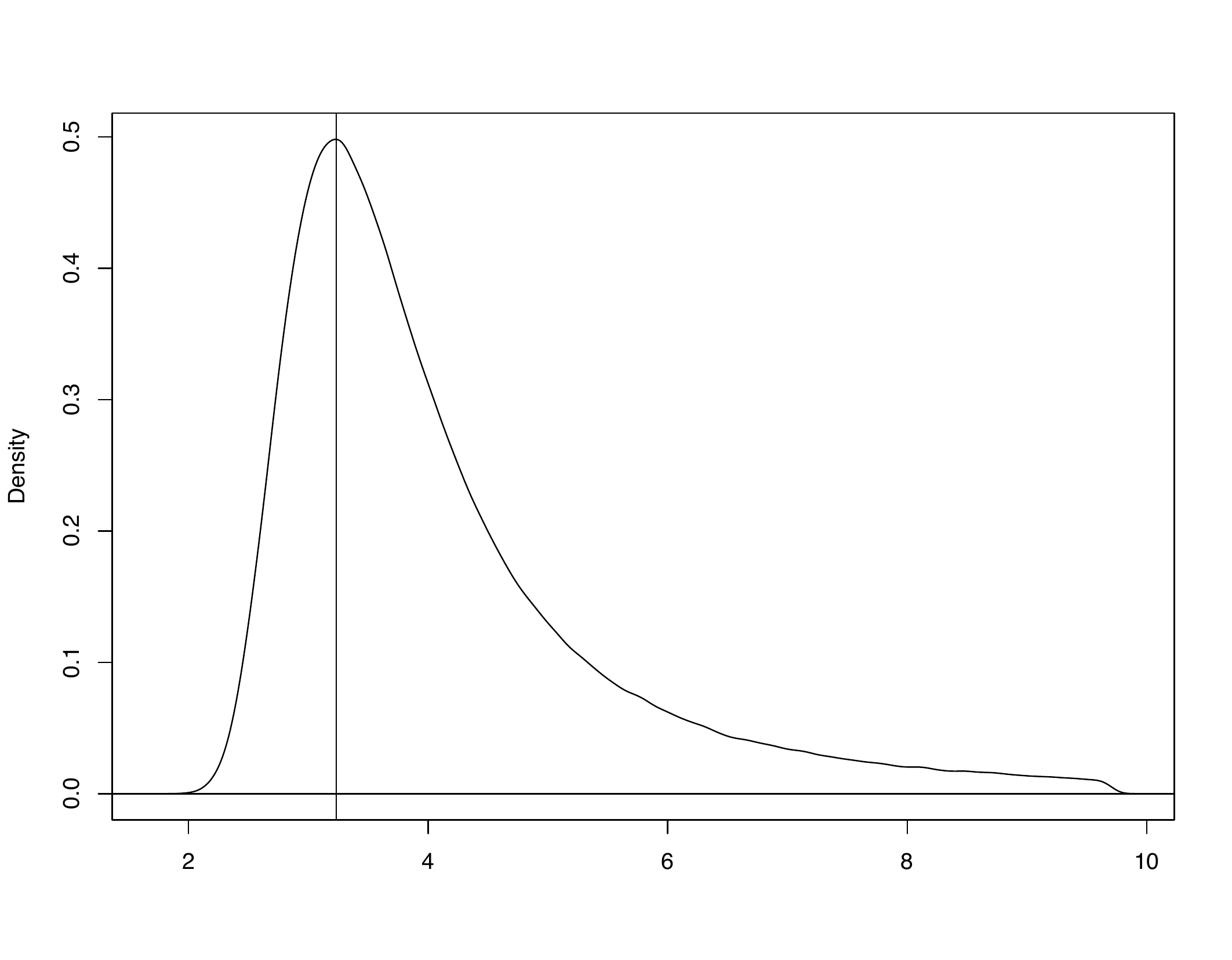}
    \end{subfigure}
     \begin{subfigure}{.26\textwidth}
        \centering
        $k(\nu=3, T=500)$
        \includegraphics[width=\linewidth]{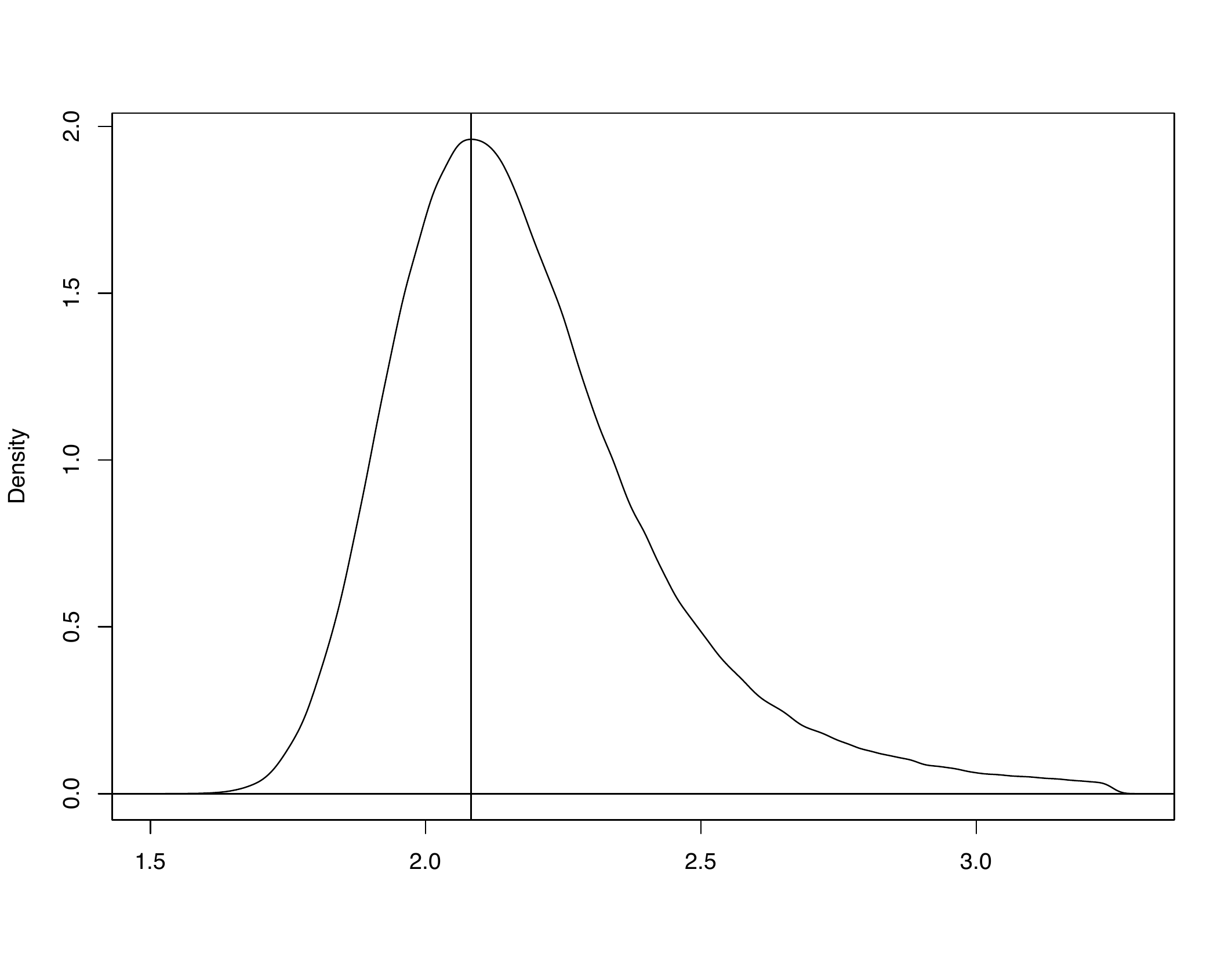}
    \end{subfigure}
    \begin{subfigure}{.26\textwidth}
        \centering
        $k(\nu=50, T=500)$
        \includegraphics[width=\linewidth]{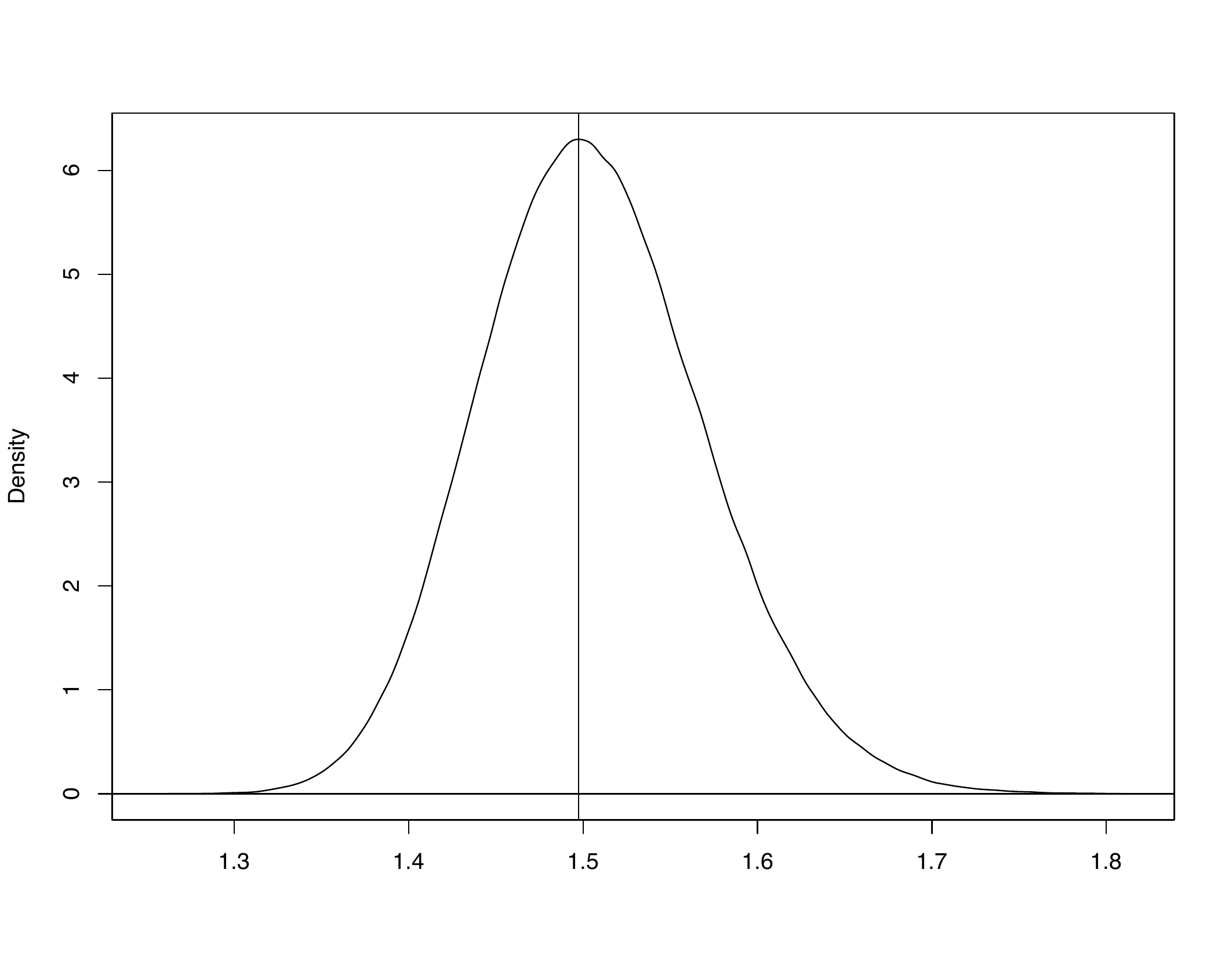}
    \end{subfigure}
      \begin{subfigure}{.26\textwidth}
        \centering
        $k(\nu=1000, T=500)$
        \includegraphics[width=\linewidth]{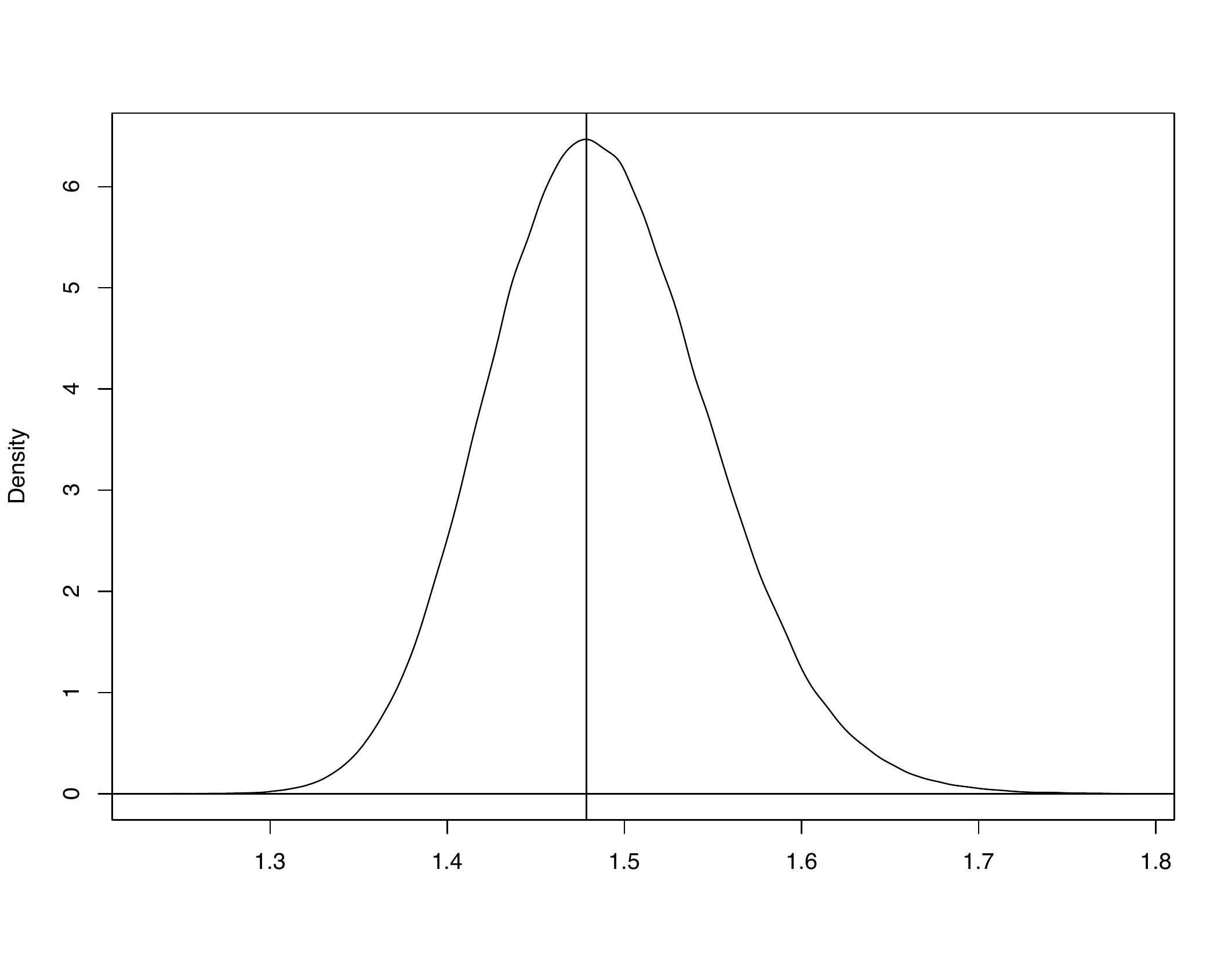}
    \end{subfigure}
    \caption{ Empirical density functions of $k$ obtained under the assumption that the error process follows as a generalized Student's $t$-distribution. %%%%%%%%%%%controlla
    $N=700000$ replications, $T=500$ and $\nu_{0}$=(1.2, 1.4, 1.8, 3, 50, 1000) are considered.}
\end{figure}

\begin{figure}[H]
\centering
    \begin{subfigure}{.26\textwidth}
        \centering
        $k(\nu=1.5, T=100)$
        \includegraphics[width=\linewidth]{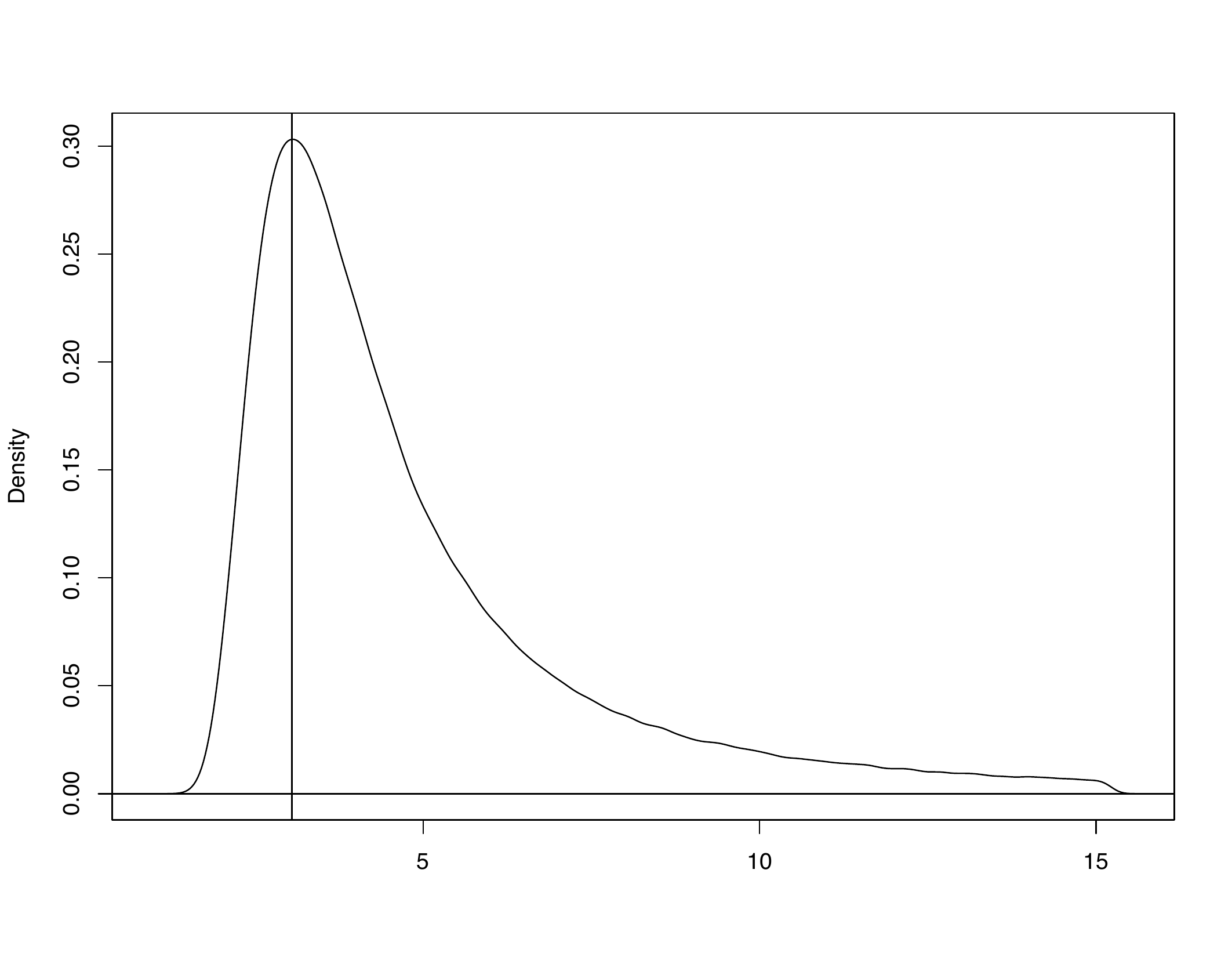}
    \end{subfigure}
    \begin{subfigure}{.26\textwidth}
        \centering
        $k(\nu=1.5, T=200)$
        \includegraphics[width=\linewidth]{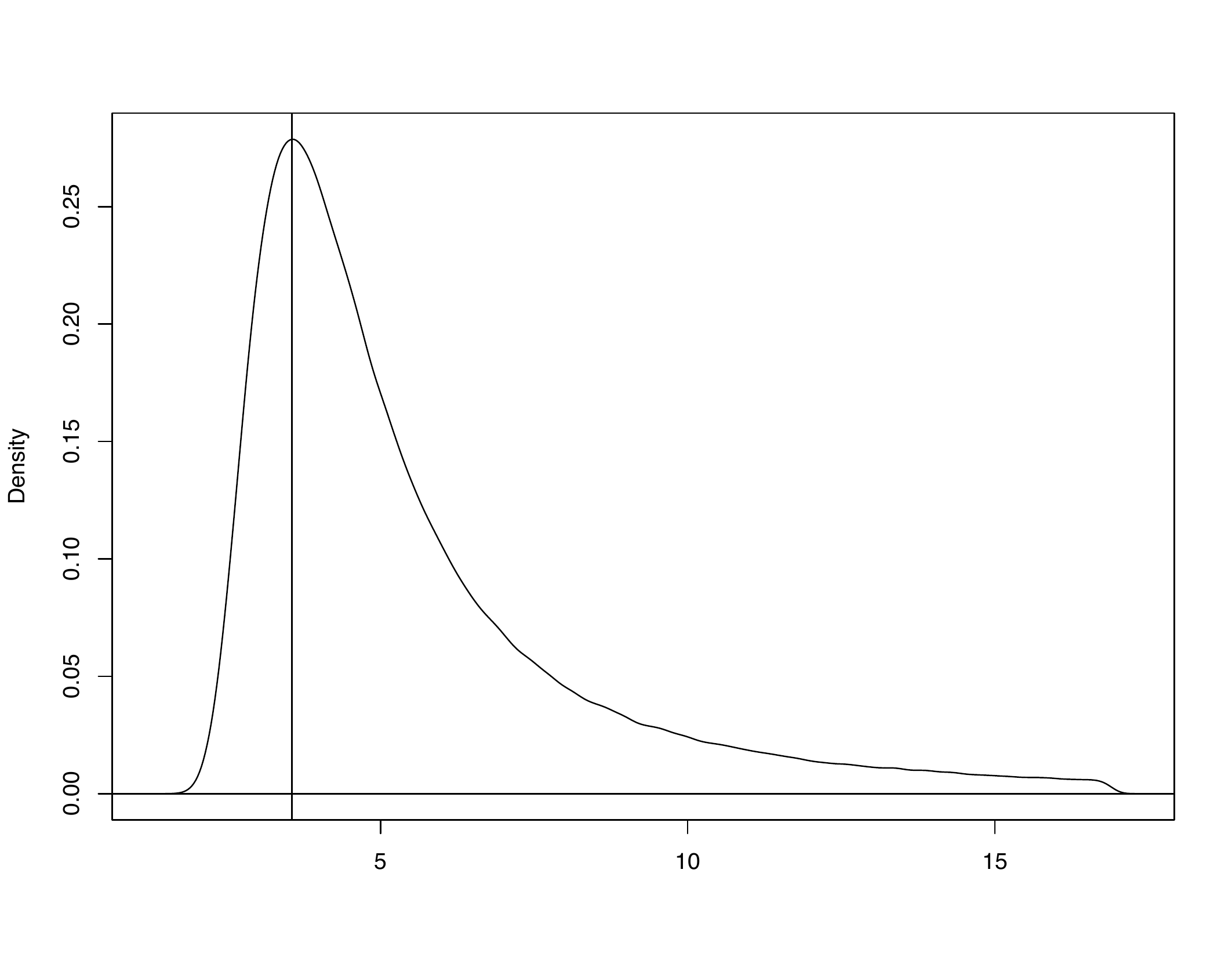}
    \end{subfigure}
       \begin{subfigure}{.26\textwidth}
        \centering
        $k(\nu=1.5, T=500)$
        \includegraphics[width=\linewidth]{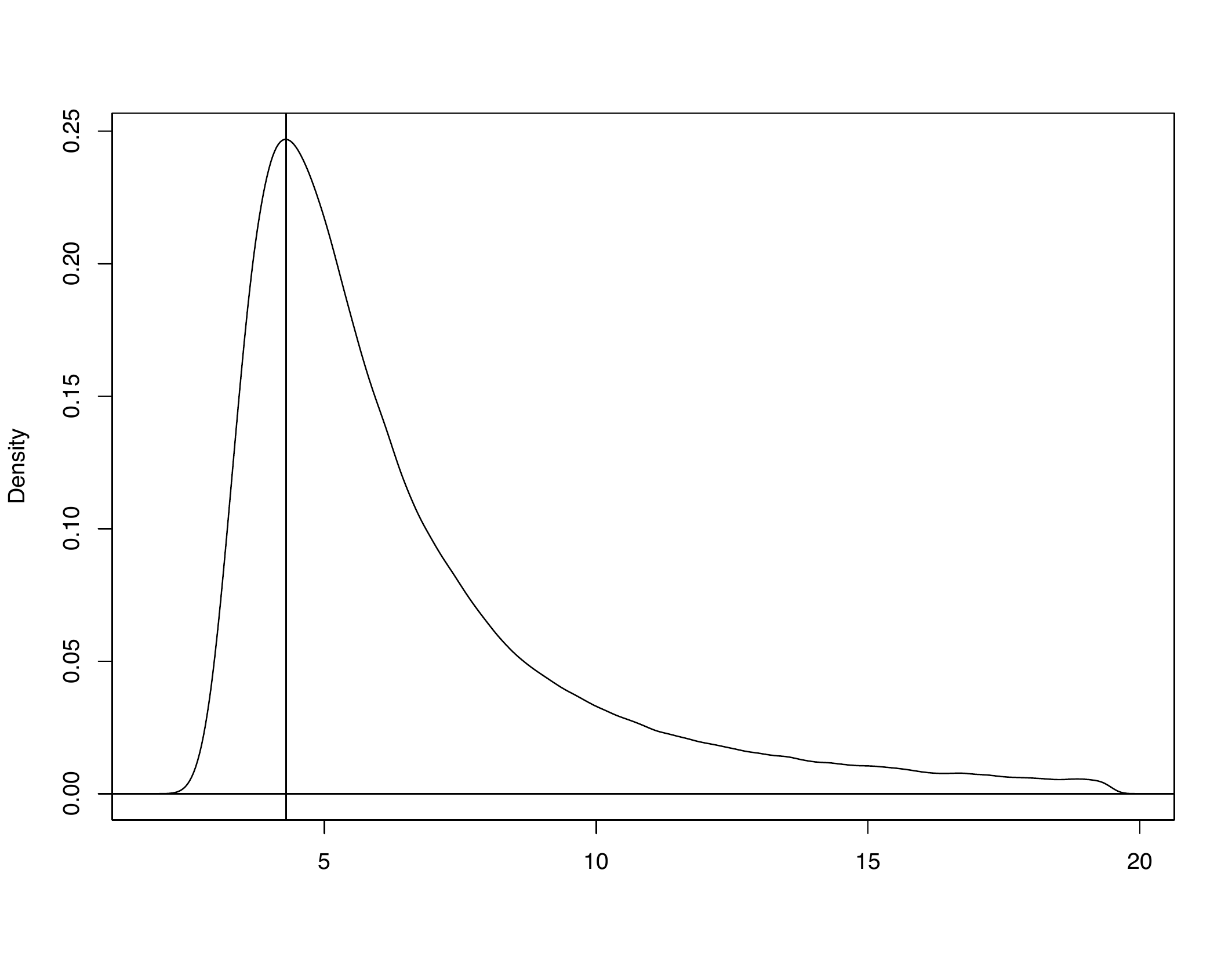}
    \end{subfigure}
    \begin{subfigure}{.26\textwidth}
        \centering
        $k(\nu=1.5, T=1000)$
        \includegraphics[width=\linewidth]{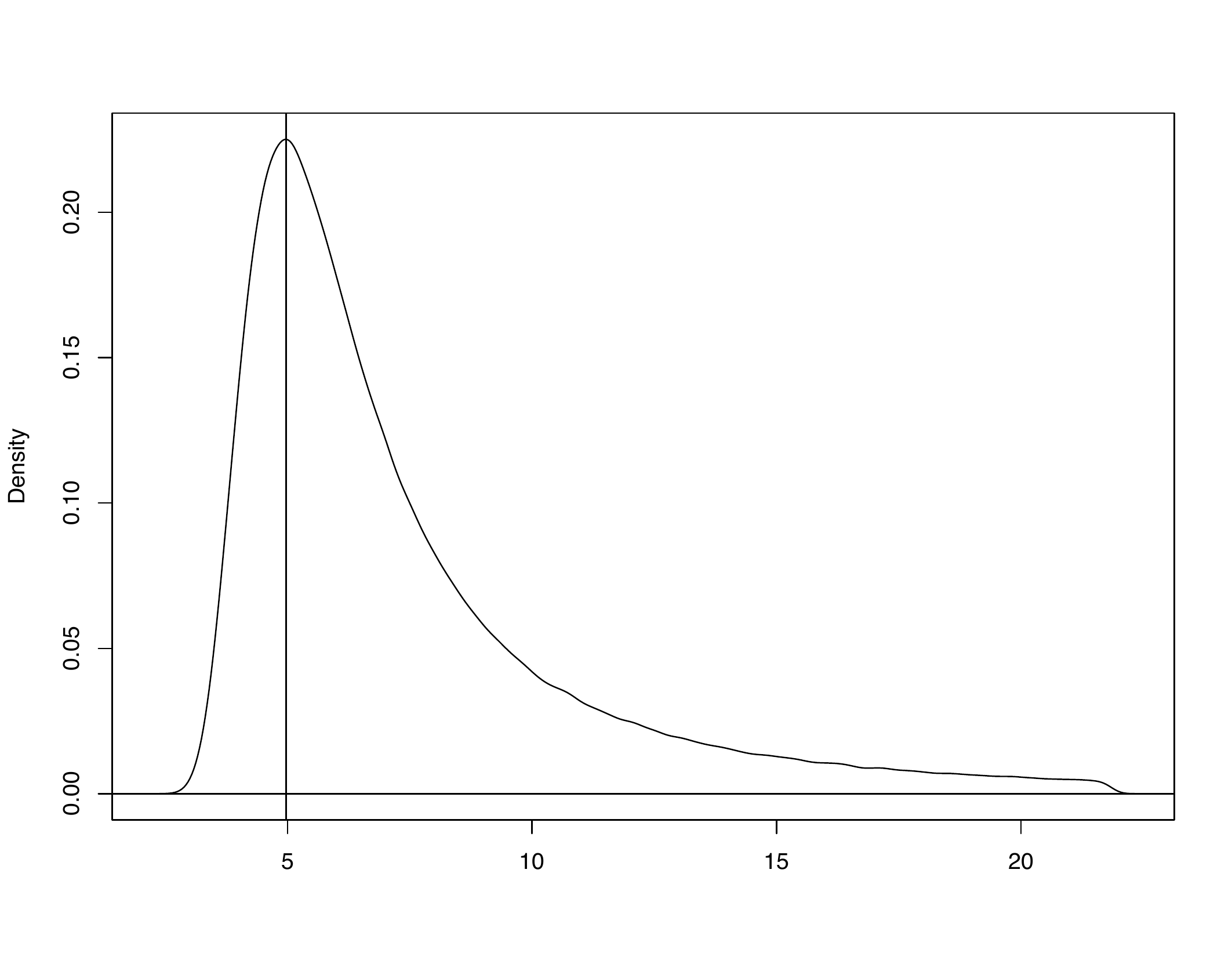}
    \end{subfigure}
      \begin{subfigure}{.26\textwidth}
        \centering
        $k(\nu=1.5, T=3000)$
        \includegraphics[width=\linewidth]{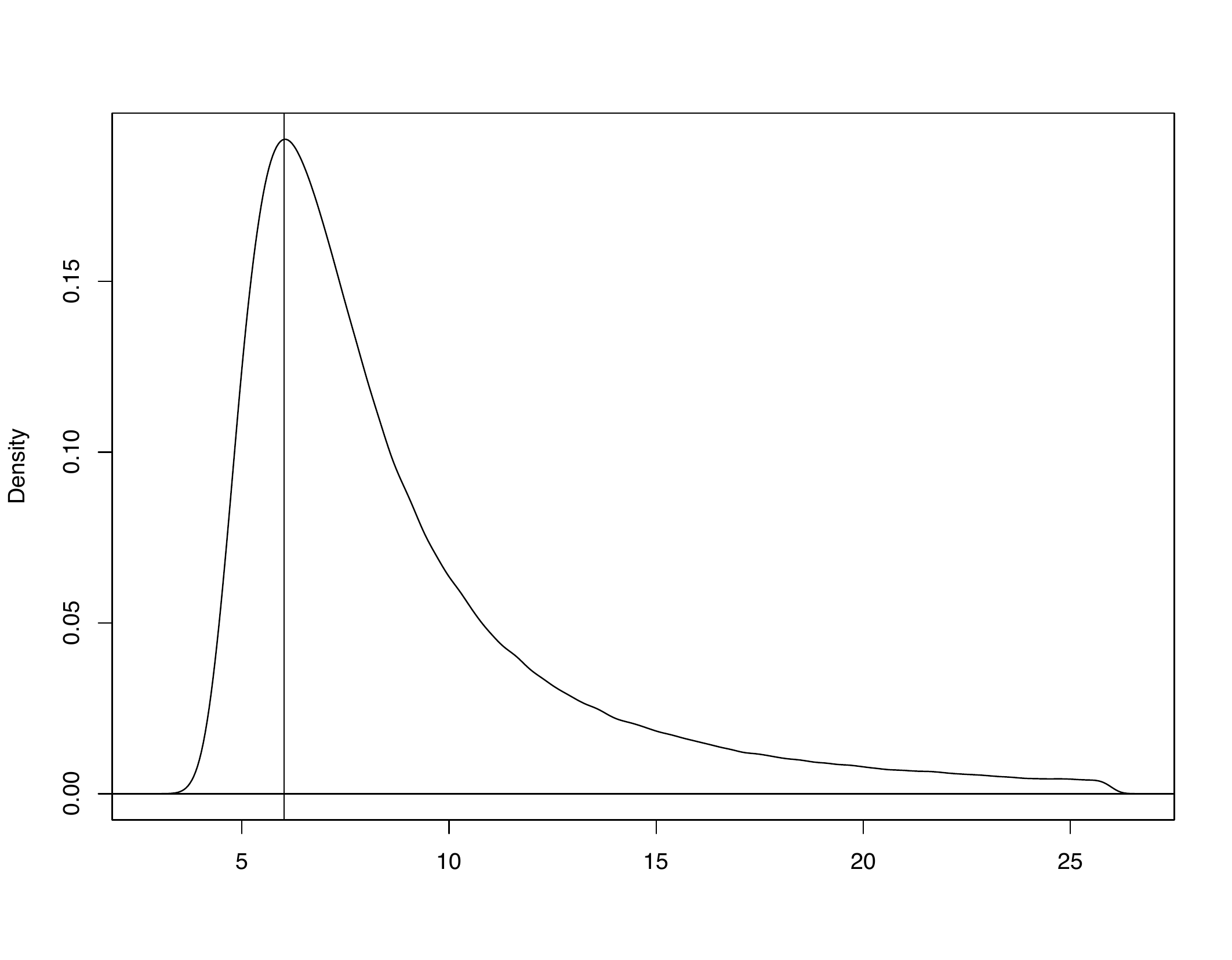}
    \end{subfigure}
     \caption{Empirical density functions of $k$ obtained under the assumption that the error process follows as a generalized Student's $t$-distribution.
       $N=700000$ replications, $T=(100, 200, 500, 1000, 3000)$ and $\nu=1.5$ are considered.}
\end{figure}

\section{Monte Carlo simulations}
Let us now analyze the numerical stability of our new robust estimator of $\sigma_{\hat{\epsilon}}$ as sample sizes increase. In other words, 
we want to analyze how our new estimator behaves as $T$ rises. To this end, we run several Monte Carlo simulations characterized by $N=10000$ 
replications each. The data generating process is a MAR(1,1) with a scale parameter of $\eta_{0}=3$ and sample sizes of $T=(100, 200, 500, 1000, 2000, 3000)$. 
We also consider several degrees of freedom $\nu_{0}=(1.2, 1.5, 1.8, 3)$ and various combinations of causal and noncausal coefficients, that is:
\begin{itemize}
\item $\phi_{0}$=0, $\varphi_{0}$=0;
\item $\phi_{0}$=0.65, $\varphi_{0}$=0.35;
\item $\phi_{0}$=0.5, $\varphi_{0}$=0.5;
\item $\phi_{0}$=0.35, $\varphi_{0}$=0.65.
\end{itemize}
For each replication, we estimate the standard deviation of the residuals using equation (9). Finally, we generate different boxplots to display the results. Figures 4-5
show how the robust standard deviation of the residuals performs whenever a MAR(1,1) with $\phi_{0}=0.65$, $\varphi_{0}=0.35$ and $\phi_{0}=0.35$, $\varphi_{0}=0.65$ is considered. 
The other data generating processes are very similar in results and available upon request.
\begin{figure}[H]
\centering
MAR(1,1): $\phi_{0}=0.65, \ \varphi_{0}=0.35$\par\bigskip
    \begin{subfigure}{.28\textwidth}
        \centering
        $\nu_{0}$=1.2, $\sigma_{0}=\infty$
        \includegraphics[width=\linewidth]{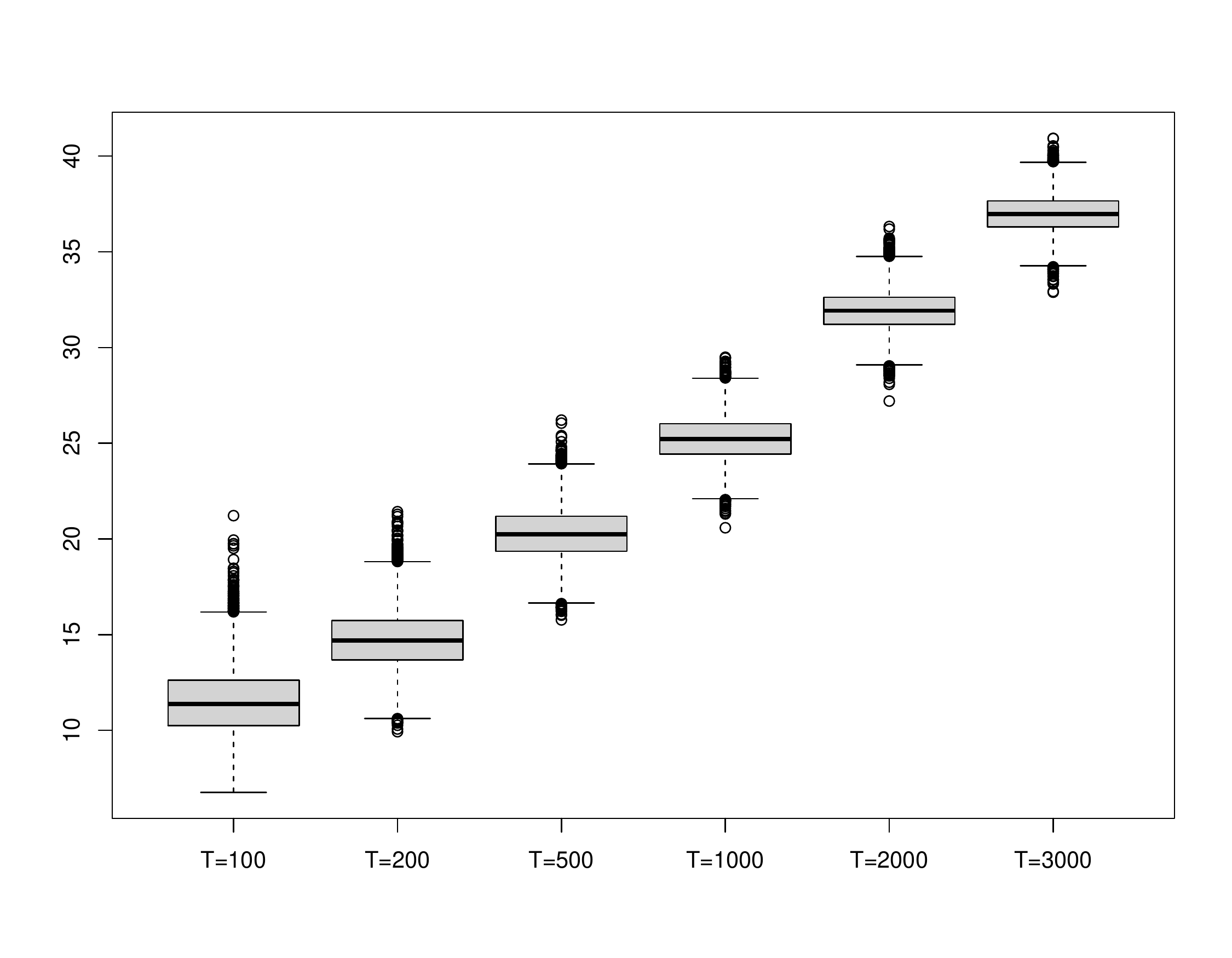}
    \end{subfigure}
        \begin{subfigure}{.28\textwidth}
        \centering
        $\nu_{0}$=1.5, $\sigma_{0}=\infty$
        \includegraphics[width=\linewidth]{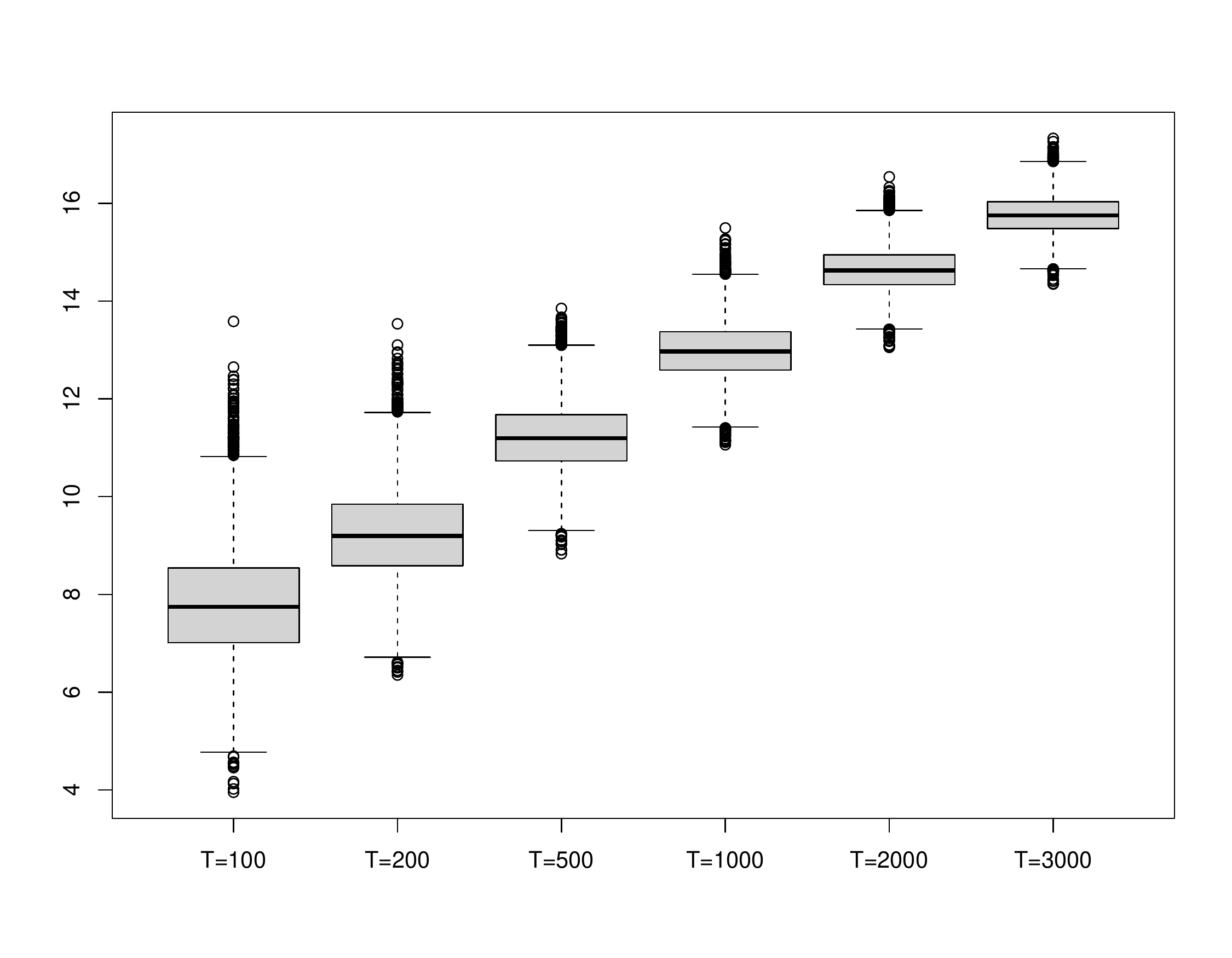}
    \end{subfigure}\\
        \begin{subfigure}{.28\textwidth}
        \centering
        $\nu_{0}$=1.8, $\sigma_{0}=\infty$
        \includegraphics[width=\linewidth]{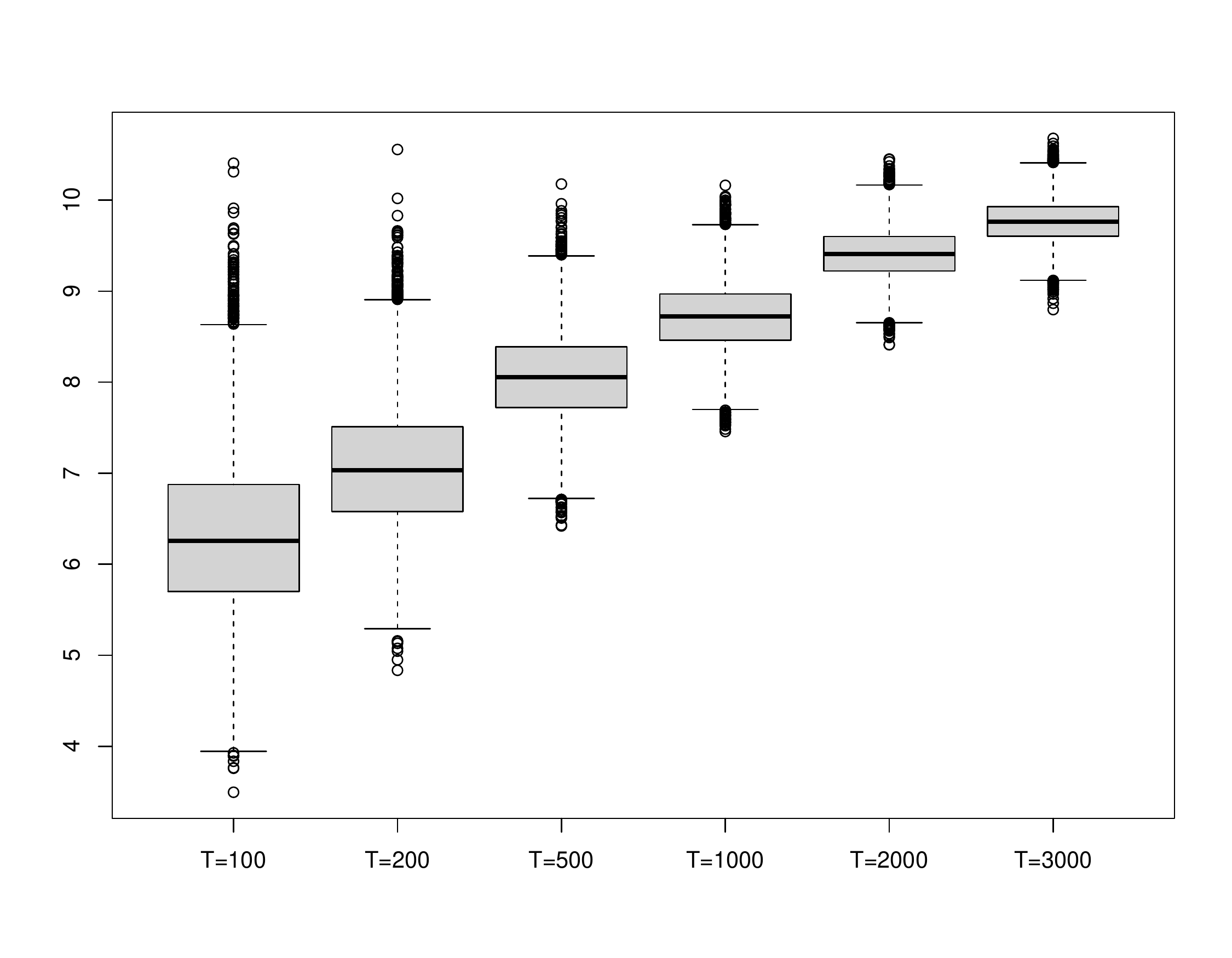}
    \end{subfigure}
    \begin{subfigure}{.28\textwidth}
        \centering
        $\nu_{0}$=3, $\sigma_{0}$=5.1
        \includegraphics[width=\linewidth]{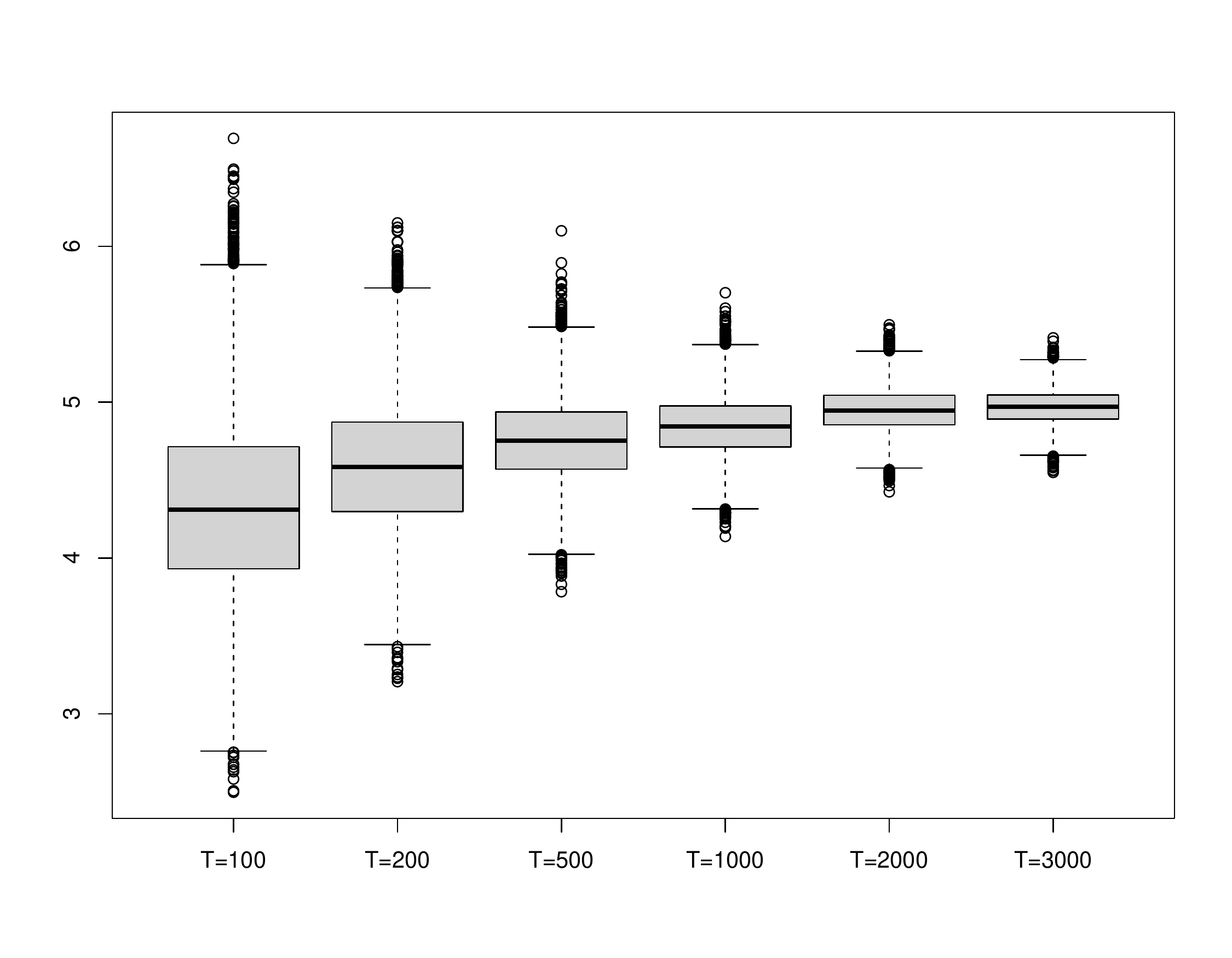}
    \end{subfigure}
	\caption{The Boxplots display how the robust estimator of the residuals behaves as $T$ increases. 
	A MAR(1,1) with $\phi_{0}=0.65$,  $\varphi_{0}=0.35$ is considered.}
\end{figure}

\begin{figure}[H]
	\centering
MAR(1,1): $\phi_{0}=0.35, \ \varphi_{0}=0.65$\par\bigskip
    \begin{subfigure}{.28\textwidth}
        \centering
        $\nu_{0}=1.2$, $\sigma_{0}=\infty$
        \includegraphics[width=\linewidth]{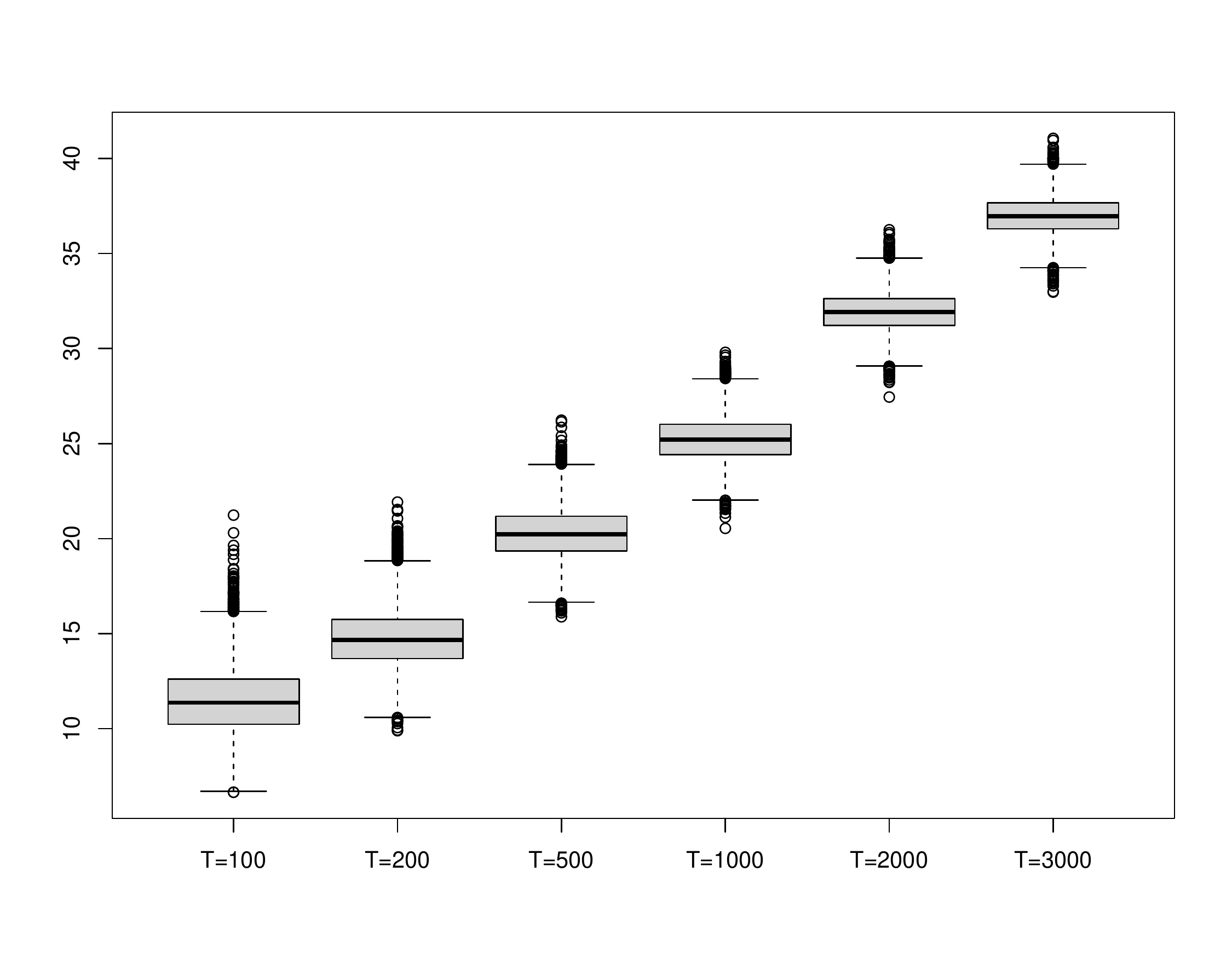}
    \end{subfigure}
        \begin{subfigure}{.28\textwidth}
        \centering
        $\nu_{0}=1.5$, $\sigma_{0}=\infty$
        \includegraphics[width=\linewidth]{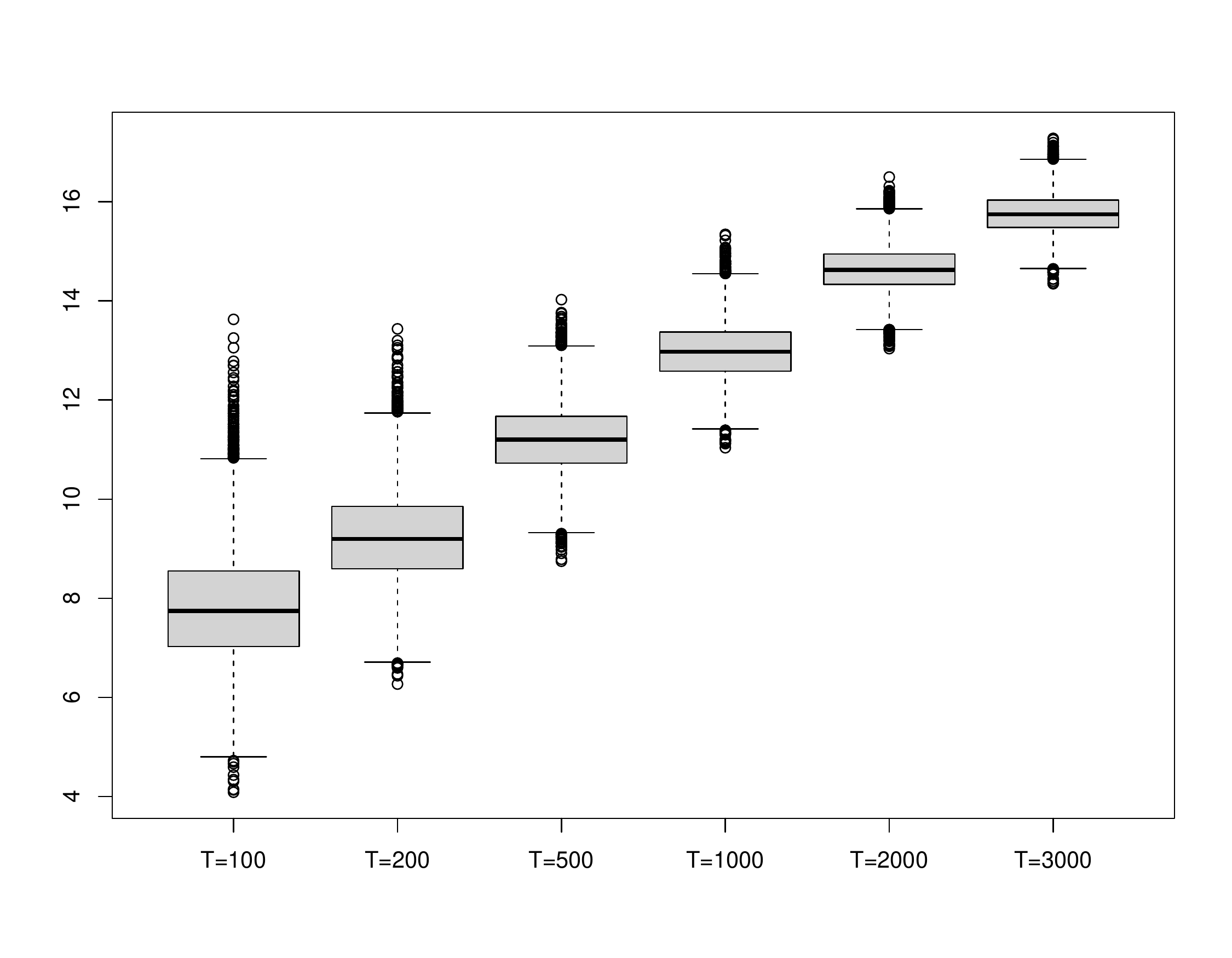}
    \end{subfigure}\\
       \begin{subfigure}{.28\textwidth}
        \centering
        $\nu_{0}=1.8$, $\sigma_{0}=\infty$
        \includegraphics[width=\linewidth]{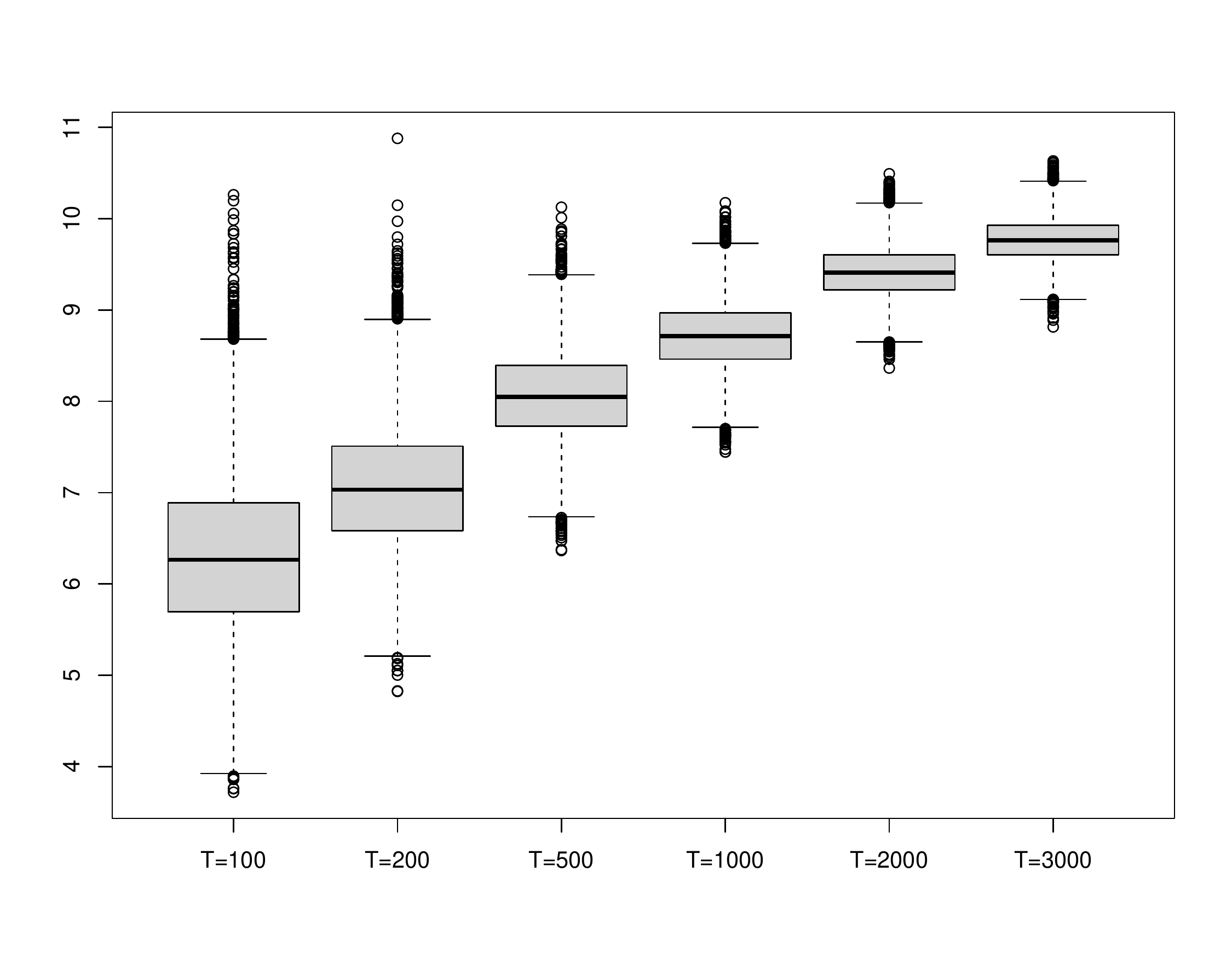}
    \end{subfigure}
    \begin{subfigure}{.28\textwidth}
        \centering
        $\nu_{0}=3$, $\sigma_{0}$=5.1
        \includegraphics[width=\linewidth]{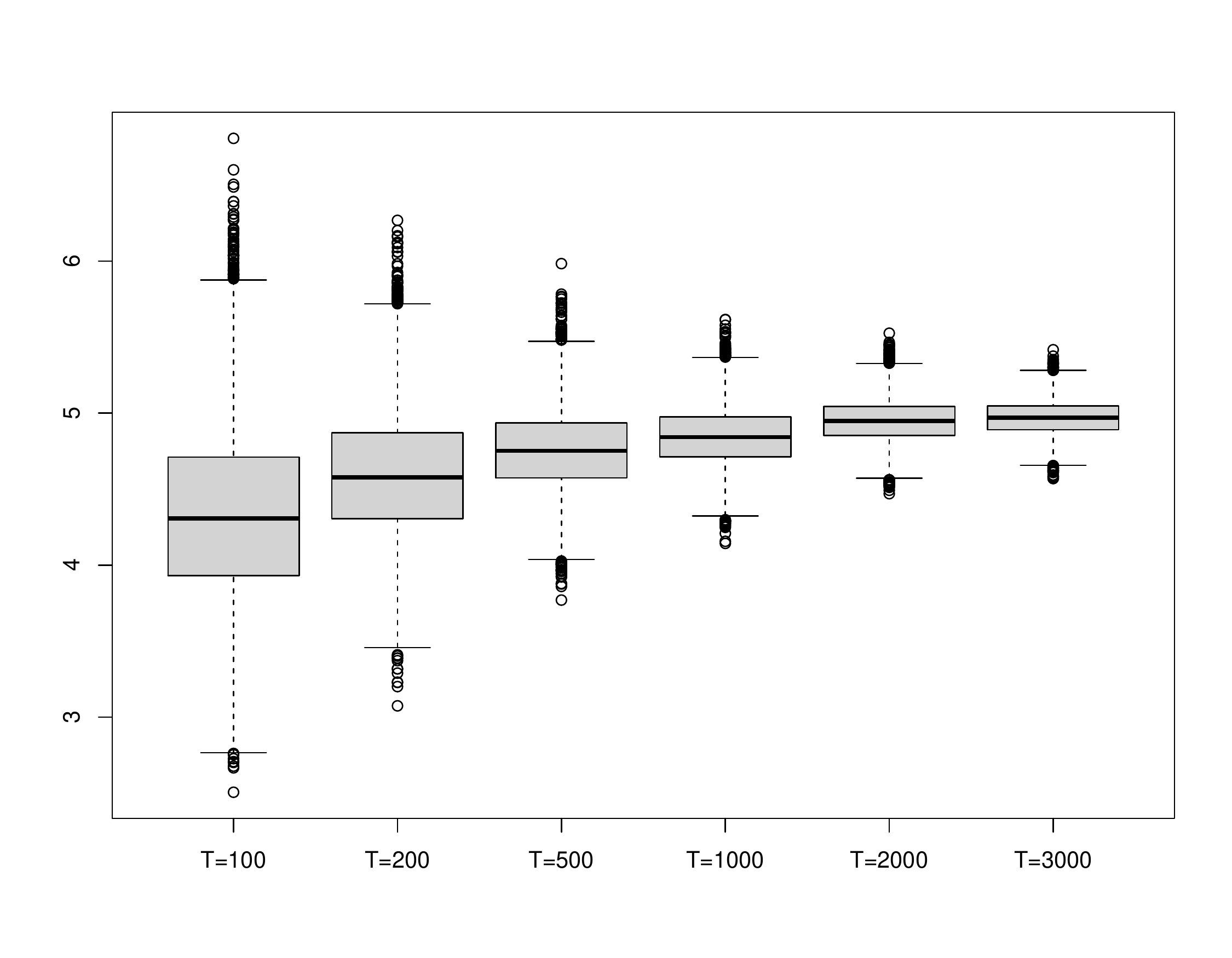}
    \end{subfigure}
    \caption{The Boxplots display how the robust estimator of the residuals behaves as $T$ increases. A MAR(1,1) with $\phi_{0}=0.35$, $\varphi_{0}=0.65$ is considered.}
\end{figure}
In the finite variance framework ($\nu_{0}=3$), the median of the estimated standard deviation goes towards its real value as the sample size increases. On the other 
hand, for $\nu_{0}=(1.2, 1.5, 1.8)$, $\sigma_{\hat{\epsilon}}$ goes to infinity regardless of which causal and noncausal coefficients are
considered. This is particularly evident in the case $\nu_{0}=1.2$, where fatter tails characterize the distribution of the residuals. However, it is important to underline 
that  the robust estimate of $\sigma_{\hat{\epsilon}}$ never explodes dramatically to infinity: in all cases considered, an increment of the sample size of $100\%$
implies an increment of the median of the estimated standard deviation of less than $100\%$. This means that the robust estimate of $\sigma_{\hat{\epsilon}}$ diverges to infinity as $T$ increases. 
However, it does so at a slow rate.\\ 
\indent To understand if the aforementioned divergence to infinity could be a problem in the inference framework, we analyze the E.R.F. of the $t$-test and compare them to the nominal 
significance level. Therefore, we run several Monte Carlo simulations with the same data generating process as before. For each replication, we test whether the estimated causal and noncausal coefficients are 
equal to their respective true values. In particular, we compute two different $t-$tests: $H_{0}$: $\phi=\phi_{0}$ and $H_{0}$: $\varphi=\varphi_{0}$ against the two-sided 
alternatives, $\phi \neq \phi_{0}$ and $\varphi \neq \varphi_{0}$ respectively. Tables 1-3 show the empirical rejection frequencies obtained at the nominal significance level of 5$\%$. The frequencies appear when the standard errors are computed using the new methodology and those introduced in Section 2. These tables do not display all the results obtained by our data generating process as they are all similar, and including them is unnecessary. 
The results obtained by MAR(1,1) with $\phi_{0}=0.35$, $\varphi_{0}=0.65$ and $\nu_{0}=1.5$ are available on request. We observe that for $\nu>2$ (Table 1), the method developed by \cite{hecq2016identification} 
provides an empirical $t-$distribution characterized by tails fatter than a standard normal distribution.
The reason for this is that in the denominator of the $t-$test, the standard errors are underestimated. This is true because of the assumption of block diagonality made on the matrix (8) 
(see Section 2). We also observe that our new approach generates fewer distortions for small sample sizes ($T = (100, 200)$) than those provided by $\Sigma$. Indeed, the E.R.F 
obtained by the matrix (6) only get closer to the $5\%$ nominal rejection frequency for $T=1000$. On the other hand, for $\nu \leq 2$ (Tables 2-3), the matrix $\Sigma$ cannot be derived. 
Furthermore, the E.R.F. generated by $\Sigma_{D}$ are still far from the nominal significance level (especially for small sample sizes and small values of $\nu$). 
Our new approach ($\Sigma_{R}$) is unique in that it provides a $t$-distribution close to the standard normal distribution. Also, it should be noted that the empirical rejection frequencies generated 
by this methodology decrease as the size of the sample becomes larger. This is due to the fact that within this framework the robust estimator of $\sigma_{\hat{\epsilon}}$ diverges to infinity as $T$ increases. However, 
as previously stated, since the divergence to infinity occurs slowly, the convergence towards 0 of the empirical rejection frequencies is so low that it is negligible in finite sample sizes.
\begin{table}[H]
\caption{Percentage of observations outside the interval [-1.96, +1.96] considering different combinations of causal and noncausal coefficients and 3 degrees of freedom. This value is equal to 5\% in a standard normal distribution.}
\resizebox{\columnwidth}{!}{
\begin{tabular}{l c c c c c c c c c}
\toprule
\multicolumn{10}{c}{Empirical rejection frequencies - nominal ones at 5\%; MAR(1,1): $\phi_{0}=0, \ \varphi_{0}=0, \ \nu_{0}=3$} \\
\toprule
Sample size & $\hat{\Sigma}$  &&   $\hat{\Sigma}_{D}$ && $\hat{\Sigma}_{R}$ & \\
\toprule
&	$\phi$ \ \ \ \ \ \  \  \ \ \ $\varphi$     &&  $\phi$ \ \ \ \ \ \  \  \ \ \  $\varphi$   &&  $\phi$ \ \ \ \ \ \  \  \ \ \  $\varphi$\\
\toprule
T=100 	&24.26\% \ \ \ \ \ \  \ 23.64\%	&&18.36\% \ \ \ \ \ \  \ 18.45\%	&&9.52\% \ \ \ \ \ \  \ 9.44\% \\ 
T=200	&14.93\% \ \ \ \ \ \  \ 15.43\%	&&14.61\% \ \ \ \ \ \  \ 15.33\%	&&7.21\% \ \ \ \ \ \  \ 7.53\% \\
T=500	&8.98\% \ \ \ \ \ \  \ 9.35\%		&&12.14\%  \ \ \ \ \ \  \ 12.25\%	&&5.26\% \ \ \ \ \ \  \ 5.76\% \\
T=1000 	&7.17\% \ \ \ \ \ \  \ 7.28\%		&&10.50\% \ \ \ \ \ \  \ 10.73\%	&&4.74\% \ \ \ \ \ \  \ 4.56\% \\
T=2000	&6.96\% \ \ \ \ \ \  \ 7.05\%		&&10.58\%  \ \ \ \ \ \  \ 10.46\%	&&5.03\% \ \ \ \ \ \  \ 5.00\% \\
T=3000 	&6.40\% \ \ \ \ \ \  \ 6.29\%		&&9.70\% \ \ \ \ \ \  \ 9.89\%	&&4.73\% \ \ \ \ \ \  \ 4.65\% \\
\toprule
\multicolumn{10}{c}{ MAR(1,1): $\phi_{0}=0.65, \ \varphi_{0}=0.35, \ \nu_{0}=3$}\\
\toprule
Sample size & $\hat{\Sigma}$  &&   $\hat{\Sigma}_{D}$ && $\hat{\Sigma}_{R}$ & \\
\toprule
&	$\phi$ \ \ \ \ \ \  \  \ \ \ $\varphi$     &&  $\phi$ \ \ \ \ \ \  \  \ \ \  $\varphi$   &&  $\phi$ \ \ \ \ \ \  \  \ \ \  $\varphi$\\
\toprule
T=100	&22.37\% \ \ \ \ \ \ 22.36\%			&&16.85\% \ \ \ \ \ \  \ 16.53\%		&&8.21\% \ \ \ \ \ \ 8.20\% \\ 
T=200 	&13.91\%  \ \ \ \ \ \ 13.90\%	       	&&12.55\% \ \ \ \ \ \  \ 13.31\%		&&4.47\% \ \ \ \ \ \ 4.80\% \\
T=500      	&8.22\%  \ \ \ \ \ \ \ 8.42\%	       	 	&&10.26\% \ \ \ \ \ \  \ 10.27\%		&&4.76\% \ \ \ \ \ \ 5.16\% \\
T=1000    	&7.38\%  \ \ \ \ \ \ \ 6.95\%		    	&&9.66\% \ \ \ \ \ \  \ 9.37\% 		&&4.82\% \ \ \ \ \ \ 4.27\% \\
T=2000	&6.42\% \ \ \ \ \ \  \ 6.93\%			&&9.33\% \ \ \ \ \ \  \ 9.64\%		&&4.68\% \ \ \ \ \ \ 4.88\% \\
T=3000 	&6.47\% \ \ \ \ \ \  \ 6.56\%			&&9.45\% \ \ \ \ \ \  \ 9.14\%		&&4.74\%  \ \ \ \ \ \ 4.77\% \\
\toprule
\multicolumn{10}{c}{MAR(1,1): $\phi_{0}=0.5, \ \varphi_{0}=0.5, \ \nu_{0}=3$}\\
\toprule
Sample size & $\hat{\Sigma}$  &&   $\hat{\Sigma}_{D}$ && $\hat{\Sigma}_{R}$ & \\
\toprule
&	$\phi$ \ \ \ \ \ \  \  \ \ \ $\varphi$     &&  $\phi$ \ \ \ \ \ \  \  \ \ \  $\varphi$   &&  $\phi$ \ \ \ \ \ \  \  \ \ \  $\varphi$\\
\toprule
T=100      	&23.50\% \ \ \ \ \ \ 23.80\%		&&17.44\% \ \ \ \ \ \ 17.35\%	&&8.53\% \ \ \ \ \ \ 8.58\% \\ 
T=200  	&14.60\%  \ \ \ \ \ \ 14.97\%	&&14.26\% \ \ \ \ \ \  14.65\%	&&5.19\% \ \ \ \ \ \ 5.21\% \\
T=500      	&8.82\%  \ \ \ \ \ \ \ 8.95\%		&&11.55\% \ \ \ \ \ \  11.91\%	&&5.08\% \ \ \ \ \ \ 5.36\% \\
T=1000    	&7.30\%  \ \ \ \ \ \ \ 7.04\%		 &&10.64\% \ \ \ \ \ \  10.25\%     &&4.54\% \ \ \ \ \ \ 4.44\% \\
T=2000	&6.70\% \ \ \ \ \ \  \ 6.81\%		&&10.39\% \ \ \ \ \ \  10.28\%	&&4.90\% \ \ \ \ \ \ 4.98\% \\
T=3000 	&6.30\% \ \ \ \ \ \  \ 6.16\%		&&10.12\% \ \ \ \ \ \  \ 9.91\% \	&&4.60\% \ \ \ \ \ \ 4.37\% \\
\bottomrule
\end{tabular}}
\caption*{The first column, $\hat{\Sigma}$, indicates the empirical rejection frequencies obtained by the matrix defined in (6). The last two columns  $\hat{\Sigma_{D}}$ and $\hat{\Sigma_{R}}$ indicate the empirical rejection frequencies obtained using the standard errors developed by Hecq et al. (2016) and our new robust approach respectively.}
\end{table}

\begin{table}[H]
\caption{Percentage of observations outside the interval [-1.96, +1.96] considering different combinations of causal and noncausal coefficients and 1.8 degrees of freedom. This value is equal to 5\% in a standard normal distribution.}
\resizebox{\columnwidth}{!}{
\begin{tabular}{l c c c c c c c c c}
\toprule
\multicolumn{10}{c}{E.R.F. - nominal ones at 5\%, MAR(1,1): $\phi_{0}=0, \ \varphi_{0}=0, \ \nu_{0}=1.8$}\\
\toprule
Sample size & $\hat{\Sigma}$  &   $\hat{\Sigma}_{D}$ & $\hat{\Sigma}_{R}$ & \\
\toprule
&	$\phi$ \ \ \ \ \ \  \  \ \ \ $\varphi$     &  $\phi$ \ \ \ \ \ \  \  \ \ \  $\varphi$   &  $\phi$ \ \ \ \ \ \  \  \ \ \  $\varphi$ &\\
\toprule
T=100  	&\ / \ \ \ \ \ \ \ \ \ \ / \ \    		&12.20\% \ \ \ \ \ 12.63\%		&6.18\% \ \ \ \ \ \ 6.41\% &\\ 
T=200 	&\ / \ \ \ \ \ \ \ \ \ \ / \ \  	   		&\ 9.62\% \ \ \ \ \ \ 10.21\%		&5.18\% \ \ \ \ \ \ 5.73\% &\\
T=500    	&\ / \ \ \ \ \ \ \ \ \ \ / \ \ 	     		&8.73\% \ \ \ \ \ \ 8.67\%		&4.71\% \ \ \ \ \ \ 4.74\% &\\
T=1000    	&\ / \ \ \ \ \ \ \ \ \ \ / \ \  			&7.88\% \ \ \ \ \ \ 8.68\%		&4.18\% \ \ \ \ \ \ 4.89\% &\\
T=2000	&\ / \ \ \ \ \ \ \ \ \ \ / \ \ 			&8.02\% \ \ \ \ \ \ 7.85\%		&4.73\% \ \ \ \ \ \ 4.54\% &\\
T=3000 	&\ / \ \ \ \ \ \ \ \ \ \ / \ \ 	 		&7.78\% \ \ \ \ \ \ 7.62\%		&4.25\% \ \ \ \ \ \ 4.16\% &\\
\toprule
\multicolumn{10}{c}{
\centering
MAR(1,1): $\phi_{0}=0.65, \ \varphi_{0}=0.35, \ \nu_{0}=1.8$}\\
\toprule
Sample size & $\hat{\Sigma}$  &   $\hat{\Sigma}_{D}$ & $\hat{\Sigma}_{R}$ & \\
\toprule
& 	$\phi$ \ \ \ \ \ \  \  \ \ \ $\varphi$      &  $\phi$ \ \ \ \ \ \  \  \ \ \  $\varphi$   &  $\phi$ \ \ \ \ \ \  \  \ \ \  $\varphi$ &\\
\toprule
T=100 	&\ / \ \ \ \ \ \ \ \ \ \ / \ \		&10.93\% \ \ \ \ \ \ 11.41\%		&5.54\% \ \ \ \ \ \ 5.73\% &\\ 
T=200	&\ / \ \ \ \ \ \ \ \ \ \ / \ \ 		&7.82\% \ \ \ \ \ \ 9.28\%	   	&4.36\% \ \ \ \ \ \ 5.04\% &\\
T=500	&\ / \ \ \ \ \ \ \ \ \ \ / \ \		&7.18\% \ \ \ \ \ \ 8.10\%		&3.74\% \ \ \ \ \ \ 4.56\% &\\
T=1000	&\ / \ \ \ \ \ \ \ \ \ \ / \ \ 	       	&6.81\% \ \ \ \ \ \ 8.32\%      	&3.77\% \ \ \ \ \ \ 4.45\% &\\
T=2000	&\ / \ \ \ \ \ \ \ \ \ \ / \ \ 		&6.51\% \ \ \ \ \ \ 7.35\%		&3.97\% \ \ \ \ \ \ 4.24\% &\\
T=3000 	&\ / \ \ \ \ \ \ \ \ \ \ / \ \ 		&6.72\% \ \ \ \ \ \ 7.23\%		&3.77\% \ \ \ \ \ \ 3.80\% &\\
\toprule
\multicolumn{10}{c}{MAR(1,1): $\phi_{0}=0.5, \ \varphi_{0}=0.5, \ \nu_{0}=1.8$}\\
\toprule
Sample size & $\hat{\Sigma}$  &   $\hat{\Sigma}_{D}$ & $\hat{\Sigma}_{R}$ & \\
\toprule
&	$\phi$ \ \ \ \ \ \  \  \ \ \ $\varphi$       &  $\phi$ \ \ \ \ \ \  \  \ \ \  $\varphi$   &  $\phi$ \ \ \ \ \ \  \  \ \ \  $\varphi$ &\\
\toprule
T=100      	&\ / \ \ \ \ \ \ \ \ \ \ / \ \			&11.72\% \ \ \ \ \ \ 11.54\%		&5.58\% \ \ \ \ \ \ 6.08\% &\\ 
T=200  	&\ / \ \ \ \ \ \ \ \ \ \ / \ \		      	&8.81\% \ \ \ \ \ \ 9.14\%	     	&4.58\% \ \ \ \ \ \ 4.74\% &\\
T=500      	&\ / \ \ \ \ \ \ \ \ \ \ / \ \			&7.63\% \ \ \ \ \ \ 7.67\%	     	&3.99\% \ \ \ \ \ \ 4.08\% &\\
T=1000    	&\ / \ \ \ \ \ \ \ \ \ \ / \ \ 		      	&7.34\% \ \ \ \ \ \ 7.96\%     	&4.04\% \ \ \ \ \ \ 4.41\% &\\
T=2000	&\ / \ \ \ \ \ \ \ \ \ \ / \ \ 			&7.11\% \ \ \ \ \ \ 7.10\%		&4.04\% \ \ \ \ \ \ 4.09\% &\\
T=3000 	&\ / \ \ \ \ \ \ \ \ \ \ / \ \ 			&7.07\% \ \ \ \ \ \ 6.76\%		&3.97\% \ \ \ \ \ \ 3.76\%&\\
\bottomrule
\end{tabular}}
\end{table}

\begin{table}[H]
\caption{Percentage of observations outside the interval [-1.96, +1.96] considering different combinations of causal and noncausal coefficients and 1.2 degrees of freedom. This value is equal to 5\% in a standard normal distribution.}
\resizebox{\columnwidth}{!}{
\begin{tabular}{l c c c c c c c c c}
\toprule
\multicolumn{10}{c}{E.R.F. - nominal ones at 5\%; MAR(1,1): $\phi_{0}=0, \ \varphi_{0}=0, \ \nu_{0}=1.2$} \\
\toprule
Sample size & $\hat{\Sigma}$  &   $\hat{\Sigma}_{D}$ & $\hat{\Sigma}_{R}$ & \\
\toprule
&	$\phi$ \ \ \ \ \ \  \  \ \ \ $\varphi$       &  $\phi$ \ \ \ \ \ \  \  \ \ \  $\varphi$   &  $\phi$ \ \ \ \ \ \  \  \ \ \  $\varphi$ &\\
\toprule
T=100      	&\ / \ \ \ \ \ \ \ \ \ \ / \ \	   	&14.91\% \ \ \ \ \ \ 14.34\%	    	 &6.11\% \ \ \ \ \ \ 6.16\%& \\ 
T=200  	&\ / \ \ \ \ \ \ \ \ \ \ / \ \ 	         &13.85\% \ \ \ \ \ \ 14.08\%	 &5.91\% \ \ \ \ \ \ 6.24\% &\\
T=500      	&\ / \ \ \ \ \ \ \ \ \ \ / \ \	         &13.34\% \ \ \ \ \ \ 12.93\%	 &6.03\% \ \ \ \ \ \ 5.28\% &\\
T=1000    	&\ / \ \ \ \ \ \ \ \ \ \ / \ \           &12.95\% \ \ \ \ \ \ 12.89\%         &5.62\% \ \ \ \ \ \ 5.56\%& \\
T=2000	&\ / \ \ \ \ \ \ \ \ \ \ / \ \ 		&12.75\% \ \ \ \ \ \ 12.74\%		&5.56\% \ \ \ \ \ \ 5.49\% &\\
T=3000 	&\ / \ \ \ \ \ \ \ \ \ \ / \ \ 		&12.26\% \ \ \ \ \ \ 12.48\%		&5.42\% \ \ \ \ \ \ 5.14\% &\\
\toprule
\multicolumn{10}{c}{MAR(1,1): $\phi_{0}=0.65, \ \varphi_{0}=0.35, \ \nu_{0}=1.2$}\\
\toprule
Sample size & $\hat{\Sigma}$  &   $\hat{\Sigma}_{D}$ & $\hat{\Sigma}_{R}$ & \\
\toprule
& 	$\phi$ \ \ \ \ \ \  \  \ \ \ $\varphi$        &  $\phi$ \ \ \ \ \ \  \  \ \ \  $\varphi$   &  $\phi$ \ \ \ \ \ \  \  \ \ \  $\varphi$ &\\
\toprule
T=100      &\ / \ \ \ \ \ \ \ \ \ \ / \ \			&12.30\% \ \ \ \ \ 13.39\%		&4.62\% \ \ \ \ \ \ 5.69\% &\\ 
T=200  	&\ / \ \ \ \ \ \ \ \ \ \ / \ \ 			&10.69\% \ \ \ \ \ \ 13.02\%		&4.51\% \ \ \ \ \ \ 5.59\% &\\
T=500      &\ / \ \ \ \ \ \ \ \ \ \ / \ \			&\ 9.22\% \ \ \ \ \ \  11.57\%	&4.04\% \ \ \ \ \ \ 4.26\% &\\
T=1000    &\ / \ \ \ \ \ \ \ \ \ \ / \ \ 			&\ 8.61\% \ \ \ \ \ \  11.38\%        &3.53\% \ \ \ \ \ \ 4.81\% &\\
T=2000	&\ / \ \ \ \ \ \ \ \ \ \ / \ \ 			&\ 8.81\% \ \ \ \ \ \  11.42\%	&3.84\% \ \ \ \ \ \ 4.74\% &\\
T=3000 	&\ / \ \ \ \ \ \ \ \ \ \ / \ \ 			&\ 8.78\% \ \ \ \ \ \  10.70\%	&3.78\% \ \ \ \ \ \ 4.58\% &\\
\toprule
\multicolumn{10}{c}{MAR(1,1): $\phi_{0}=0.5, \ \varphi_{0}=0.5, \ \nu_{0}=1.2$}\\
\toprule
Sample size & $\hat{\Sigma}$  &   $\hat{\Sigma}_{D}$ & $\hat{\Sigma}_{R}$ & \\
\toprule
&	$\phi$ \ \ \ \ \ \  \  \ \ \ $\varphi$     &  $\phi$ \ \ \ \ \ \  \  \ \ \  $\varphi$   &  $\phi$ \ \ \ \ \ \  \  \ \ \  $\varphi$ &\\
\toprule
T=100      	&\ / \ \ \ \ \ \ \ \ \ \ / \ \	              	&13.26\% \ \ \ \ \ \ 12.65\% 	&5.31\% \ \ \ \ \ \ 5.09\%	&\\ 
T=200  	&\ / \ \ \ \ \ \ \ \ \ \ / \ \ 	  	     	&12.12\% \ \ \ \ \ \ 12.00\%		&4.90\% \ \ \ \ \ \ 4.91\% 	&\\
T=500      	&\ / \ \ \ \ \ \ \ \ \ \ / \ \	                 &10.67\% \ \ \ \ \ \ 10.72\%		&4.51\% \ \ \ \ \ \ 3.98\%	&\\
T=1000    	&\ / \ \ \ \ \ \ \ \ \ \ / \ \                   &\ 9.82\% \ \ \ \ \ \ 10.35\% 		&3.86\% \ \ \ \ \ \ 4.12\% 	&\\
T=2000	&\ / \ \ \ \ \ \ \ \ \ \ / \ \ 		     	&10.27\% \ \ \ \ \ \ 10.36\%		&4.22\% \ \ \ \ \ \ 4.00\% 	&\\
T=3000 	&\ / \ \ \ \ \ \ \ \ \ \ / \ \ 		     	&9.61\% \ \ \ \ \ \ 9.99\%		&4.27\% \ \ \ \ \ \ 4.44\% 	&\\
\bottomrule
\end{tabular}}
\end{table}

\section{Empirical investigations}
We illustrate the differences and similarities in the computed standard errors of MAR models on three 
time series. These are (a) the monthly wheat prices from January 1990 until September 2020 (source: IMF),
(b) the monthly inflation rate in Brazil, obtained from year to year difference on the IPCA index (the IPCA targets population 
families with household income ranging from 1 to 40 minimum wages and this income range guarantees a 90$\%$
coverage of families living in 13 geographic zones) observed from January 1997 to June 2020 (source: Central Bank of Brazil) and (c)
the variation of daily COVID-19 deaths in Belgium from 10/March/2020 to 17/July/2020 (source: WHO).
Figure 6 presents the data. With this panel of applications, we want to show that MAR models are also interesting 
for modeling other series as well as the usual commodity prices.
\begin{figure}[H]
	\begin{center}
 		 \begin{subfigure}[t]{.4\textwidth}
   		 	\centering
  		 	\includegraphics[width=\linewidth]{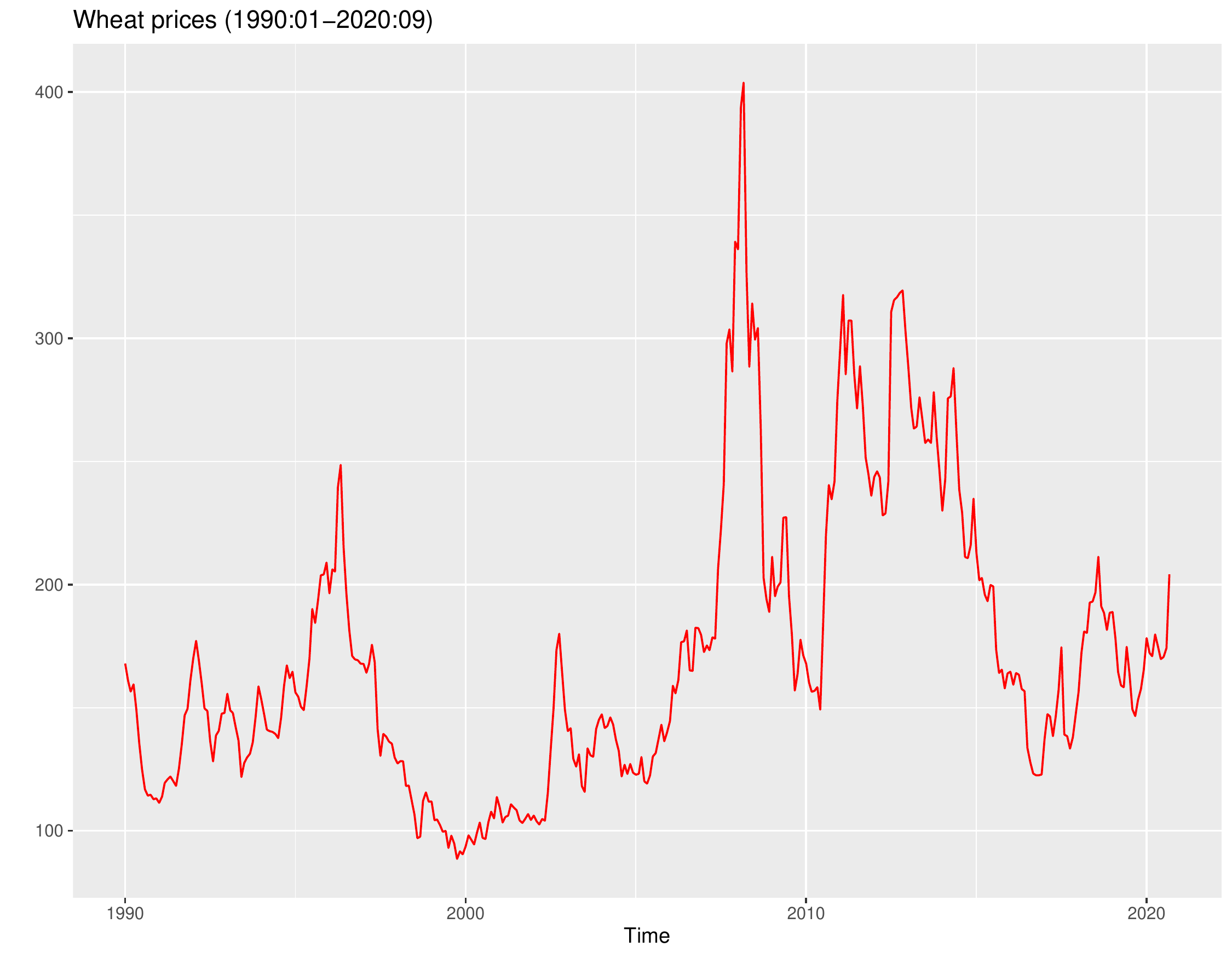}
 		  	\caption{\textit{Monthly data for the wheat prices.}}
 		 \end{subfigure}
		\begin{subfigure}[t]{.4\textwidth}
   		 	\centering
   			 \includegraphics[width=\linewidth]{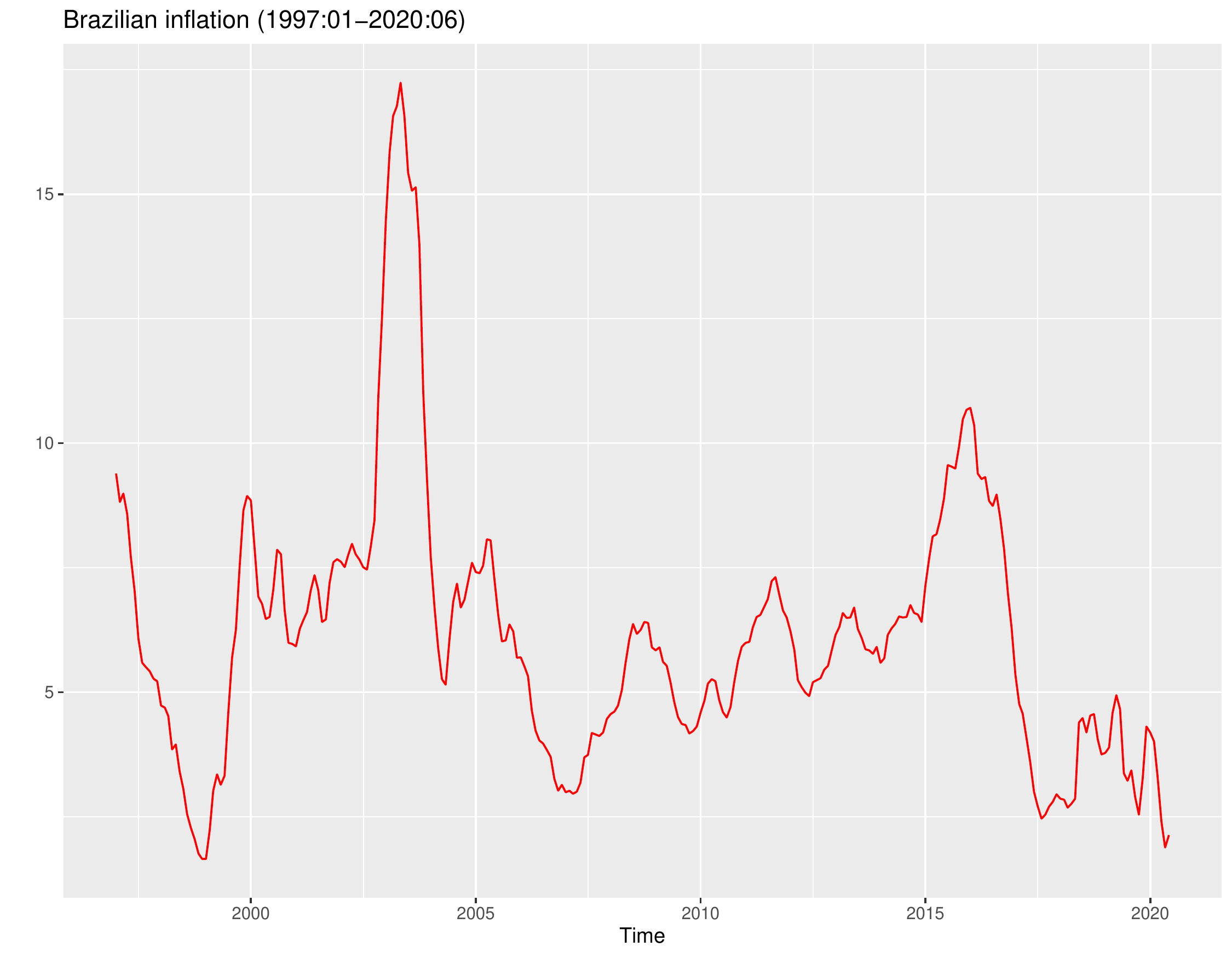}
  		 	\caption{\textit{Monthly data for the inflation rate in Brazil.}}
 	\end{subfigure}

	\end{center}
\medskip
     	\begin{center}
		\begin{subfigure}[t]{.4\textwidth}
   			 \centering
   		 	 \includegraphics[width=\linewidth]{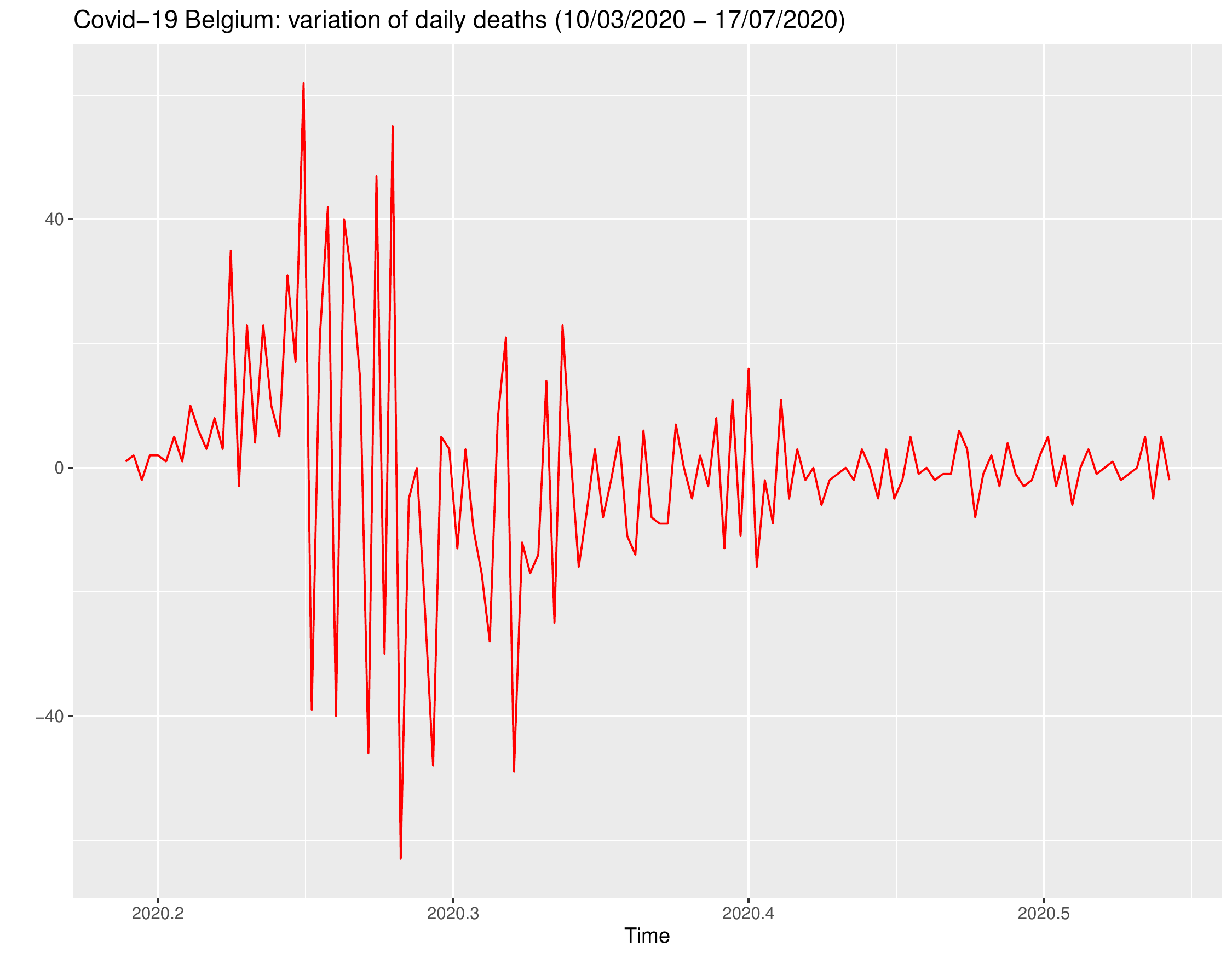}
   		 	\caption{\textit{Daily data for the variation of deaths for COVID-19 in Belgium.}}
 		\end{subfigure}
	 \end{center}
 \caption{Charts of the 3 time series covered by the empirical investigation.}
 \end{figure}
Estimating MAR models involves carrying out a series of steps. Firstly, it is necessary to estimate a conventional causal autoregressive model by %cambia covid 19
OLS in order to obtain the lag order $p$ using information criteria (see \cite{lanne2011noncausal}, and \cite{hecq2016identification}). 
We find $p = 2$ for the inflation rate and wheat prices, whereas $p = 4$ is chosen for Belgian’s COVID-19 series. Secondly, using an AML approach 
and searching for the $r$ and $s$ (with $p=r+s$) that maximize the generalized Student’s $t$ likelihood function, we discover that 
wheat prices and Brazilian inflation follow a MAR(1,1). On the other hand, the variation of COVID-19 deaths follows a MAR(2,2). 
Finally, we detail the value of estimated parameters and their standard errors using the methods reviewed and the novel method 
introduced in this paper.\\
\indent The results of the simulations carried out lead us to expect some differences and similarities given the degrees of freedom estimated for the three variables: 
for COVID-19 data $\hat{\nu}=1.17$, on wheat prices $\hat{\nu}=2.21$ and $\hat{\nu}=3.22$ for Brazilian inflation. Although we observe fat tails 
in each series, only Belgian daily data is characterized by degrees of freedom below 2. However, none of them are significantly different from 2. 
To verify this, we use the standard errors given by $-(T-p)^{-1}\delta^{2}l_{T}(\hat{\boldsymbol{\phi}}, \hat{\boldsymbol{\varphi}},\boldsymbol{\theta_{2}}) / \delta \boldsymbol{\theta_{2}} \delta\boldsymbol{\theta_{2}}^{\prime}$ with $\hat{\boldsymbol{\theta_{2}}}=(\hat{\nu}, \hat{\eta})$, which is a consistent estimator of the expected Fisher information matrix of the distributional parameters 
(see \cite{lanne2011noncausal}). Unlike $\Sigma$, $\Omega$ has no restrictions and can also be computed when the population variance is infinite.\\
\indent Differences in the standard errors of the causal and noncausal parameters depend on the approaches used to compute them.  % controlla hat matrici
In the empirical application concerning the Brazilian inflation rate (Table 5), we notice that the standard errors 
obtained through the robust estimator of residuals are larger than those obtained by the "traditional" methodologies described in Section 2. 
The same is true for the causal coefficient in the wheat prices (Table 4). For the noncausal coefficient of the same empirical application, 
we obtain smaller standard errors when they are calculated by $\hat{\Sigma}$. Finally, matrix $\hat{\Sigma}$ does not provide standard errors when the time series 
related to the variation of the number of fatalities for COVID-19 in Belgium (Table 6) is considered. This is true because it is characterized by an error term with infinite variance.
On the other hand, our methodology generates standard errors that are smaller than those obtained by the matrix $\hat{\Sigma}_{D}$.
\newline

\begin{table}[H]
\caption{Estimated coefficients and standard errors for wheat prices.}
\resizebox{\columnwidth}{!}{
\begin{tabular}{l c c c c c c c c c}
\toprule
\multicolumn{10}{c}{
\centering
Wheat prices} \\
\toprule
\textbf{Estimated}   &&&&&  \textbf{Standard errors} &&&&\\
\textbf{coefficients} &&&  $\hat{\Sigma}$  &&   $\hat{\Sigma}_{D}$ && $\hat{\Sigma}_{R}$ & \\
\toprule
$\widehat{\phi_{1}}$=0.9241 		 		&&& 0.007851 	&& 0.003549	&& 	0.013969\\
$\widehat{\varphi_{1}}$=0.2866 		&&&0.019681 	&& 0.023949	&&  	0.035019\\
$\widehat{\eta}$=6.9191					&&& 0.515185 	&& 0.515185 	&& 	0.515185\\
$\widehat{\nu}$=2.2096 				&&& 0.324037 	&& 0.324037 	&& 	0.324037\\
\bottomrule
\end{tabular}}
\label{...}
\end{table}

\begin{table}[H]
\caption{Estimated coefficients and standard errors for inflation rate in Brazil.}
\resizebox{\columnwidth}{!}{
\begin{tabular}{l c c c c c c c c c}
\toprule
\multicolumn{10}{c}{
\centering
Brazilian inflation rate} \\
\toprule
\textbf{Estimated}   &&&&&  \textbf{Standard errors} &&&&\\
\textbf{coefficients} &&&  $\hat{\Sigma}$  &&   $\hat{\Sigma}_{D}$ && $\hat{\Sigma}_{R}$ & \\
 \toprule
$\widehat{\phi_{1}}$=0.5842 		 		&&& 0.038656 	&& 0.028383	&& 	0.046492\\
$\widehat{\varphi_{1}}$=0.9385 		&&& 0.016444 	&& 0.006895	&&  	0.019777\\
$\widehat{\eta}$=0.2654				&&& 0.021556 	&& 0.021556 	&& 	0.021556\\
$\widehat{\nu}$=3.2217 					&&& 0.719318 	&& 0.719318 	&& 	0.719318\\
\bottomrule
\end{tabular}}
\label{...}
\end{table}

\begin{table}[H]
\caption{Estimated coefficients and standard errors for the variation of daily COVID-19 deaths in Belgium.}
\resizebox{\columnwidth}{!}{
\begin{tabular}{l c c c c c c c c c}
\toprule
\multicolumn{10}{c}{
\centering
COVID-19 in Belgium: variation of daily deaths} \\
\toprule
\textbf{Estimated}   &&&&&  \textbf{Standard errors} &&&&\\
\textbf{coefficients} &&&  $\hat{\Sigma}$  &&   $\hat{\Sigma}_{D}$ && $\hat{\Sigma}_{R}$ & \\
\toprule
$\widehat{\phi_{1}}$=-0.4660 	 &&& / 	&&	0.028793		&& 	0.024285\\
$\widehat{\phi_{2}}$=-0.5853 	 &&&  / 	&&	0.028793		&&  	0.024277\\
$\widehat{\varphi_{1}}$=0.0803	 &&& /	&&	0.028793		&&  	0.023881\\
$\widehat{\varphi_{2}}$=0.6037	 &&&  / 	&&	 0.025619 	&&  	0.023881\\
$\widehat{\eta}=4.2279$		 &&&  0.639214	&& 0.639214 		&& 	0.639214\\
$\widehat{\nu}=1.1785$	         &&& 0.639214 	&& 0.209340 	&	& 	0.209340\\
\bottomrule
\end{tabular}}
\label{...}
\end{table}

\section{Conclusions} %controlla Exp Inf
In this paper, we review the behavior of the ML estimator for mixed causal and noncausal models. In particular, we focus on those with an error term that is assumed to be distributed according to a generalized
Student's $t-$distribution. As demonstrated by \cite{lanne2011noncausal}, the expected Fisher Information matrix of causal and noncausal parameters can be computed if and only 
if the probability density function satisfies a certain set of assumptions. Generalized Student's $t-$distributions infinite variance do not meet one of these assumptions ($\mathcal{J} >1$), hence, this methodology is 
not applicable in the context of infinite variance. This is a serious limitation since we cannot consider time series with estimated degrees of freedom equal to or less than two.
\cite{hecq2016identification} propose a new, approximate, and simplified way of computing the standard errors of these parameters. This methodology is also implemented in the R package MARX and applied in several studies. 
However, through a simulation study, we show that the aforementioned approach does not facilitate the inference of the MAR parameters. This is true because the $t$-tests exhibit empirical rejection frequencies far from 
the nominal significance level, especially when applied to small sample sizes and for small values of degrees of freedom. In order to bypass these problems, we propose an novel way to compute standard error based 
on a simple alternative estimator of the variance of residuals. Monte Carlo simulations show the optimal performance of this new estimator, even when the variance of the population is not finite. Finally, 
we estimate MAR models on three time series, and illustrate the differences in the estimated standard error produced by the various approaches discussed.

\section*{Acknowledgement}
The authors would like to thank Naiyie Lamb, Yicong Lin, Elisa Voisin, Sean Telg, Ines Wilms, two anonymous referees as well as the participants of the ICEEE 2021, Cagliari, for valuable comments and suggestions. 
All remaining errors are ours.

\bibliographystyle{elsarticle-harv}
\biboptions{authoryear}
\raggedright
\bibliography{Reviewpaper}
\section{Appendix A}
The following table shows the different values of $k$ that maximize their own empirical density functions 
according to the different values of $T$ and $\nu$.

\begin{center}
\begin{table}[H]
\scalebox{.95}{
\resizebox{\columnwidth}{!}{
\begin{tabular}{l c c c c c c c c c c c}
\toprule
\multicolumn{12}{c}{$\boldsymbol{k^{*}}$} \\
\toprule
&               $\nu=1.2$     &&$\nu=1.4$     &&$\nu=1.5$     &&$\nu=1.6$     &&$\nu=1.8$     &&$\nu=3$  \\
\toprule
T=100      &4.186322    &&3.317155     &&3.049654    &&2.866044    &&2.57295     &&1.937395  \\ 
T=200 	&5.311298    &&3.901011    &&3.557615     &&3.2330488   &&2.85024    &&2.02271 \\
T=500      &7.266156    &&4.941986    &&4.297126     &&3.849296    &&3.233094    &&2.082257 \\
T=1000    &9.022733    &&5.839081    &&4.971029     &&4.330869    &&3.491673    &&2.116381 \\
T=2000    &11.41613    &&6.827137    &&5.597711     &&4.750695     &&3.761855    &&2.158208\\
T=3000    &13.20991    &&7.448153    &&6.022052     &&5.128269    &&3.902047    && 2.166739\\
\bottomrule
\end{tabular}}}
\end{table}
\end{center}
\end{document}